\documentclass[aps,groupedaddress,nofootinbib,11pt]{revtex4}

\usepackage[dvips]{epsfig}
\usepackage{amsfonts}
\usepackage{amsmath}
\usepackage{}

\newcommand{\Eqref}[1]{(\ref{#1})}
\newcommand{\ket}[1] {\mbox{$ \vert #1 \rangle $}}
\newcommand{\kett}[1] {\mbox{$ \vert #1 \rangle_2 $}}
\newcommand{\bra}[1] {\mbox{$ \langle #1 \vert $}}
\newcommand{\abs}[1] {\mbox{$ \vert #1 \vert $}}

\newcommand{\inv}[1]{\frac{1}{#1}}

\newcommand{\av}[1]{\langle\!\langle \,  #1 \, \rangle\!\rangle}

\newcounter{subequation}[equation] \makeatletter
\expandafter\let\expandafter\reset@font\csname reset@font\endcsname
\newenvironment{subeqnarray}
  {\arraycolsep1pt
    \def\@eqnnum\stepcounter##1{\stepcounter{subequation}{\reset@font\rm
      (\theequation\alph{subequation})}}\eqnarray}
  {\endeqnarray\stepcounter{equation}}
\makeatother

\newcommand{\ba}{\begin{eqnarray}}
\newcommand{\ea}{\end{eqnarray}}
\newcommand{\sba}{\begin{subeqnarray}}
\newcommand{\sea}{\end{subeqnarray}}

\def\th{\mbox{th}}
\def\ch{\mbox{ch}}
\def\sh{\mbox{sh}}

\begin{document}

\vskip 1truecm

\title{
Inflationary
spectra and partially decohered distributions}
\vskip 1truecm
\author{David Campo}
\email[]{dcampo@math.uwaterloo.ca}
 \affiliation{Department of Applied Mathematics, University of Waterloo,
 University Avenue, Waterloo, Ontario, N2L 3G1 Canada}
\author{Renaud Parentani}
\email[]{Renaud.Parentani@th.u-psud.fr}
\affiliation{Laboratoire de Physique Th\'{e}orique, CNRS UMR 8627,
B\^atiment 210,
Universit\'{e} Paris XI, 91405 Orsay Cedex, France}
\maketitle

\vskip 1truecm
\centerline{{\bf Abstract }}
\vskip 0.5truecm
 It is generally expected that decoherence processes will erase the
 quantum properties
 of the inflationary
 primordial spectra.
 However, given the weakness of gravitational interactions,
 one might end up with a distribution which is only
 partially decohered.
 Below a certain critical change, we show that the inflationary
 distribution retains quantum 
 properties.
We identify four of these: a squeezed spread
 in some direction of phase space, non-vanishing off-diagonal matrix
 elements, and two properties used in quantum optics called
 non-$P$-representability and non-separability.
 The last two are necessary conditions to violate Bell's inequalities.
 The critical value above which all these properties are lost
 is associated to the `grain' of coherent states.
 The corresponding value of the entropy
 is equal to half the maximal (thermal) value.
 Moreover it coincides with the
 entropy of the effective distribution obtained by neglecting the
 decaying modes. By considering backreaction effects, we also
 provide an upper bound for this entropy at the onset of the
 radiation dominated era.

\vskip 1truecm

\section{Introduction}

Inflation tells us that the primordial density fluctuations arise
from the amplification of vacuum fluctuations
 \cite{Mukhanov81,Staro79,H,G,B}. As a result of this amplification, the
initial vacuum state becomes a product of highly squeezed two-mode
states \cite{Grishchuk90}.
 In spite of their
 intricate character,
expectation values
are rather simple.
For instance, when evaluated on the Last Scattering Surface,
besides the small deviations from a scale invariant spectrum
\cite{LiddleLythPhysRep}, the non-trivial information delivered by
the two-point function concerns the temporal coherence of the
modes \cite{Albrecht00}. This last property follows from the fact
that its Fourier transform, i.e. the power spectrum, displays
oscillations and zeros. Since we are dealing with an ensemble, the
presence of these zeros tells us that {\it all} realizations of
the ensemble with a given wavenumber have the same temporal phase.
 In classical terms, this coherence can be easily
 enforced by neglecting the decaying mode. The residual random properties of
the distribution then only concern the amplitude of the
 growing modes.
In the high occupation number limit, these amplitudes can be
treated as stochastic
variables.

 In brief, when
considering the physics which took place near
 the recombination,
it is convenient and sufficient to neglect the decaying mode.
However this simplified description has several
drawbacks. Indeed, the settings are too restrictive to describe
distributions which result from decoherence processes (or more
generally any non-linear process) taking place in the early
universe. In particular, generic modifications of two-point
function from the standard result
cannot be parameterized in these simplified terms.
This is true even for modifications
which preserve
the isotropy and the Gaussianity of the distribution.

The aim of this paper is to analyze quantum distributions characterized by
an arbitrary level of coherence. We identify the parameter
which governs this level and determine how the properties of the
distribution depend on it.
At fixed occupation number, two-mode squeezed states
give the most coherent distribution, and thermal states the least one.
Starting with squeezed states in the high occupation number limit,
when slightly modifying this parameter,
the properties are either extremely robust,
and therefore accessible in the classical regime, or
extremely fragile.
One finds that the fragile properties are quantum mechanical in
character. They concern the squeezed spread in some direction in
phase space (also called the sub-fluctuant direction), the quantum
coherence of macroscopically different states in the perpendicular
direction, as well as two refined properties called the
non-separability \cite{Werner89} and the non-$P$-representability
\cite{Glauber69} of the distribution. We show that both the last
properties are necessary to have expectation values which can
violate Bell's inequalities. Below a critical value of the
decoherence parameter, the distributions retain quantum
properties, and can therefore violate these inequalities.

Two-mode coherent states provide an appropriate basis to perform
this analysis. The reasons are the following. First, as well
known, a coherent state provides the quantum counterpart of a
particular classical realization of an ensemble. This
correspondence is well defined
for the highly excited (amplified) modes we are dealing with,
because the spread in position and momentum is negligible with
respect to the ``displacement'' in phase space which encodes the
occupation number. Remember that the observed temperature
anisotropies of relative amplitude $10^{-5}$ fix the mean
occupation number to be of the order of $10^{100}$.

 Second, homogeneity and isotropy restrict the non-vanishing matrix
elements of the distribution to {\it two-mode} elements with
opposite wave vectors ${\bf k}$. Hence, superpositions of
 {\it two-mode} coherent states
 can characterize Gaussian, homogeneous, and
 isotropic distribution with arbitrary level of coherence.
 Two important relationships should be noticed. First, the
 level of coherence is fixed by the expectation value
 of two destruction operators with opposite
 wave vectors.
 (This expectation value
 determines the number of quanta in {\it entangled pairs}.)
Second, the residual coherence
of the distribution is related to the residual value of the decaying mode.
Therefore the `quality' of temporal coherence of the modes at horizon
re-entry can be (in principle) traced back to the quantum distribution.

 Third, two-mode coherent states allow to make contact
 with the general remark of \cite{Zurekbook,Zurek93}
according to which squeezed states rapidly decay into a
statistical mixture of coherent states when the oscillators are
weekly coupled to an environment.
When no longer neglected, small non-linearities \cite{Maldacena03}
amongst modes in inflationary models will
also induce some decoherence.
(Let us make
clear that the apparent violation of unitarity follows,
as usual, from the neglect of the
correlations among modes
 when evaluating the entropy, see footnote p. 188 in \cite{gottfried}. If
one keeps track of these non-linearities, the non-linear evolution
would give no entropy increase.) When applying the considerations
of \cite{Zurekbook,Zurek93} to two-mode squeezed states, the
distribution becomes diagonal in the basis of two-mode coherent
states, but not of one-mode coherent states. As we shall see, this
guarantees that the temporal coherence of modes is preserved while
the quantum properties are progressively erased.

The distribution which is diagonal in the two-mode coherent states
basis plays a critical role
because it separates distributions which have
kept their quantum properties from those which have lost them. The
corresponding value of the entropy
is $1/2$ the maximal (thermal) value, given the occupation
number.

To establish the {\it critical} character of this decoherence
scheme we explore the space of (partially) decohered Gaussian
distributions. When dealing with macroscopic occupation numbers,
the properties of these distributions fall into two separate
classes: they are either extremely robust or extremely sensitive
to the level of coherence. (In more dynamical terms, this would
translate into a insensitivity/sensitivity to a weak coupling to
an environment.) The entropy is found to be a sensitive variable
whereas the power spectrum is a robust one:
 the relative modifications of the latter are
$1/n$ whereas the changes of the former are in $\ln n$.
(Throughout the paper we shall evaluate these entropies exactly by
exploiting the fact that all Gaussian distributions can be
expressed in terms of thermal distributions of new oscillators,
see Appendix C.) Similarly, the residual squeezing and the quantum
coherence are both sensitive whereas the temporal coherence of the
modes is the second robust property of the distribution. In fact,
we shall see that {\it all} quantum properties are lost for a
critical value of the residual coherence of the distribution. This
critical value is that given by the {\it grain} of coherent
states.

We then exploit the sensitivity of the entropy to establish a
correspondence between the above distribution and the effective
one obtained by neglecting the decaying mode and used in numerical
codes such as CMBFAST \cite{cmbfast}. Since their entropy
coincide, and since the entropy faithfully traces the residual
coherence, this agreement tells us the quantum distributions
 which
correspond to the effective one are very `close' to the diagonal
distribution in the two-mode coherent state basis.

Finally, we also provide an upper bound for the entropy by
considering backreaction effects at the end of the inflationary
period. This upper bound is given by $3/4$ of the maximal entropy.

Having established these properties, two questions should be
confronted. The first concerns the efficiency of decoherence
processes which occurred in nature: {\it Were they powerful enough
to reach the critical decoherence level associated with coherent
states ?} This question is currently under investigation.
Preliminary results indicate that the critical value is indeed
reached thereby implying that the resulting entropy is larger than
the above bound. The second question is more
important. It concerns the impact of the decoherence level on
structure formation (i.e. non-linear evolution).
This is a difficult question but we can already acknowledge
the fact that the critical
decoherence level associated with the loss of quantum coherence is
so small that it could hardly have any impact on this evolution.
Therefore the question of the existence of residual quantum
properties seems to be of pure academic delight.

 We conclude this introduction by mentioning
 that closely related questions have been already analyzed
 in several papers, see
 \cite{MukhaBrandyProkopecg,Albrecht94,Giovanninig,Matacz94,PolStar96,KiefPolStar98a,KiefPolStar00,cpP1}.
 What we add\footnote{The present work is based
 on our former paper \cite{cpP1}.
 For completeness and to ease the reading, we
 have included
 some of its material and shall no longer refer to it.}
 is a further clarification of the matters,
the critical role played by coherent states in separating density
matrices, the upper bound of the entropy, and the algebraic
relationships between the entropy and the other quantum properties.

\section{review of the standard derivation of primordial spectra}

\subsection{Quantum distribution of two-mode states}

In this subsection we recall how the amplification of modes
of a massless field propagating in a FRW universe translates
in quantum settings in the fact that
the initial ground state
evolves into a product of
highly squeezed {two-mode states}.
Before proceeding, we remind the reader that
it has been shown that the evolution of linearized cosmological perturbations
(metric and density perturbations) reduces to the
propagation of
massless
scalar fields in a FRW spacetime \cite{MBF92}. In this article, we
shall only consider the massless scalar test field since the
transposition of the results to physical fields represents no
difficulty. Indeed, when preserving the linearity of the
evolution, the only modification concerns the late time dependence
of the modes \cite{cmbslow}.

Let us work in a flat FRW universe. The line element is:
 \ba \label{metric}
  ds^2 &=& a(\eta)^2 \left[ -d\eta^2 +  \delta_{ij}dx^{i}dx^{j} \right] \, .
 \ea
For definiteness and simplicity, we consider a
cosmological evolution which starts with an inflationary de Sitter
phase and ends
with a 
radiation dominated period. When using the conformal time $\eta$
to parameterize the evolution, the scale factor is respectively
given by
 \sba \label{aeta}
  a(\eta) &=& - \frac{1}{H(\eta-2 \eta_r)} \, , \qquad
  {\mbox{for\, \, }} -\infty < \eta < \eta_r  \, , \\
  a(\eta) &=& \frac{1}{H\eta_r^2} \, \eta  \, , \qquad
  {\mbox{for\, \, }}  \eta > \eta_r \, ,
 \sea
where $\eta_r> 0 $ 
designates  
the end of inflation.
The transition
is such that
the scale factor and the Hubble parameter are
continuous functions.
This approximation based on an instantaneous transition is
perfectly justified for modes relevant for CMB physics. Indeed,
their wave vector $k$ obeys
$k \eta_r  \sim 10^{-25}\sim e^{-60}$ when inflation lasts for
$60$ e-folds. Hence, the phase shift they would accumulate in a
more realistic smoothed out transition would be completely
negligible.

Let $\xi(\eta,{\bf x})$ be a massless test scalar field
propagating in this background metric. It is convenient to work
with the rescaled field $\phi = a \xi $ and to decompose it into
Fourier modes
 \ba
   \phi(\eta, \bold x) &=& \int\!\!d^3k \,
   \frac{e^{i \bold k \bold x}}{(2\pi)^{3/2}}
   \phi_{\bf k}(\eta) \, .
 \ea
The time dependent mode $\phi_{\bold k}$ obeys
 \ba \label{PropPhi_k}
  \partial_{\eta}^2 \phi_{\bf k}  +
  \left(k^2 - \frac{\partial_\eta^2 a}{a} \right) \phi_{\bf k} = 0 \, ,
 \ea
where $k=\abs{\bold k}$ is the norm of the conformal wave vector.

In our background solution, ${k}^2 - {\partial_\eta^2 a}/{a}$ is
negative during the de Sitter period when the wavelength is larger
that the Hubble radius. This leads to a large amplification of
$\phi_k$. In quantum terms this mode amplification translates into
spontaneous pair production characterized by correspondingly large
occupation numbers.

To obtain the final distribution of particles, one should
introduce two sets of positive frequency solutions of Eq.
\Eqref{PropPhi_k}. The {\it in} modes are defined at asymptotic
early time, and the {\it out} ones at late time. Both have unit
positive Wronskian in conformity with the usual particle
interpretation \cite{B&D}. One gets
 \sba \label{modes}
  \phi_k^{in}(\eta) &=& \frac{1}{\sqrt{2k}}
  \left(1-\frac{i}{k(\eta - 2 \eta_r)}\right)
  e^{-ik(\eta - 2 \eta_r)} \, , \quad
  \mbox{for\, \,} \eta < \eta_r   \, , \\
  \phi_k^{out}(\eta) &=&
  \frac{1}{\sqrt{2k}} e^{-ik\eta} \, ,\quad
  \mbox{for\, \, }  \eta > \eta_r  \, .
 \sea
In spite of the time dependence of the background, these two
positive frequency modes are unambiguously defined (up to an
arbitrary constant phase which drops in all expectation values and
which has here been chosen so as to simplify the forthcoming
expressions). In the de Sitter epoch, there is no ambiguity for
relevant modes if inflation lasts more than $70$ e-folds, see
\cite{NPC} for the evaluation of the small corrections one obtains
when imposing positive frequency at some finite early time.
Moreover, there is no ambiguity for the initial state of relevant
modes: at the onset of inflation they must be in their ground
state \cite{MBF92,cmbParenta}.

In the radiation dominated era there is no ambiguity either
because the conformal frequency is constant since $\partial^2_\eta
a = 0$. In this era, the decaying and the growing modes
correspond, respectively, to the real and the imaginary part of
the out modes Eq. (\ref{modes}b),
 as can be seen by considering the limit $k \eta \ll 1$.
 (When considering only the leading order terms (in $1/k \eta_r$)
 of expectation values,
 several definitions of the growing modes can be used
  and lead to the same results. However
  in order to obtain unambiguous answers to the questions
  we shall ask (which concern next to leading order terms),
  we must
  adopt a precise definition.
  In the sequel, we define the growing mode
  by the imaginary part of $\phi_k^{out}$.
  Such a definition is necessary
  to relate, for instance,
   the power of the decaying mode to that of
   the so-called subfluctuant mode.)

The {\it in} and {\it out} modes are related by a Bogoliubov transformation
\ba \label{inoutmodes}
 \phi_k^{in}(\eta) = \alpha_k \phi_k^{out}(\eta) +
             \beta_k^* \phi_k^{out\, *}(\eta) \, ,
\ea
where the Bogoliubov coefficients are given by the Wronskians
\ba \label{abcoef}
  \alpha_{k} = \left( \phi_{k}^{out} , \phi_{k}^{in} \right)
  \mbox{, \ \ }
  \beta_{k}^{*} = -  \left( \phi_{k}^{out\, *} , \phi_{k}^{in} \right) \, .
\ea
These overlaps should be evaluated at transition time $\eta_r$
since modes satisfy different equations in each era.
One gets
 \sba \label{bogolincosmo}
   \alpha_k &=& - \frac{ e^{2i k \eta_r}}{2k^2\eta_r^2}
   \left( 1 - 2 i{k\eta_r}- 2k^2\eta_r^2 \right)
   = \frac{-1}{2k^2\eta_r^2} \, ( 1 + O(k\eta_r)^3) \, , \\
   \beta_k^{*} &=&
   \frac{1}{2k^2\eta_r^2} \, .
 \sea
Thus, for relevant modes, $k\eta_r \sim 10^{-25}$, the in modes
are enormously amplified. Concomitantly, they are dominated by the
sine during the radiation dominated era because the phase of
$-\beta_k^{*} / \alpha_k$ is much smaller than one:
 \ba \label{neglectdecay}
  \phi_k^{in}(\eta)
   &=& \frac{i}{k^{2}\eta_r^2} \left[ \frac{\sin k\eta}{\sqrt{2 k}} \,
       + \, O(k\eta_r)^3 \,
       \frac{\cos k\eta}{\sqrt{2k}}\right]
       \, , \qquad  \eta \geq \eta_r \, .
 \ea
To leading order, i.e. neglecting the cosine in Eq.
\Eqref{neglectdecay}, the physical modes $\phi^{in}_k/a$
correspond to the growing modes. They possess a well-defined
temporal behavior, e.g. they are constant until they start
oscillating as they re-enter the Hubble radius when $k \eta \simeq
1$.

Lets now see how these considerations translate in second
quantized settings. Each mode operator is decomposed twice
 \ba \label{hatphik}
  \hat \phi_{\bold k}(\eta) =
  \hat a_{\bold k}^j  \phi_{k}^j(\eta) +
  \hat a_{-\bold k}^{j\, \dagger} \phi_{k}^{j \, *}(\eta) \, ,
 \ea
where $j$ stands for both the in and out basis.
The operators so defined are related by the
transformation
 \ba \label{inoutoperator}
   \hat a_{\bold k}^{in} = \alpha_k^{*} \, \hat a_{\bold k}^{out} -
              \beta_k \, \hat a_{-\bold k}^{out \, \dagger} \, .
 \ea
This transformation
couples ${\bf k}$ to $-{\bf k}$ only.
Hence, when starting from the in vacuum (the state annihilated by the
$\hat a_{\bold k}^{in}$ operators),
every out particle
of momentum ${\bf k}$ will be accompanied by a partner
of momentum $-{\bf k}$. Moreover, pairs characterized
by different momenta are incoherent
(in the sense that in expectation values any
product of annihilation and creation operators of different momenta
will factorize).

These two properties are explicit when writing the in vacuum in
terms of out states (i.e. states with a definite out particle
content). From Eq. \Eqref{inoutoperator}, one gets (see
 \cite{MukhaBrandyProkopecg,PhysRep95})
 \ba \label{inoutvac}
  \ket{0,\, in}
  &=&
      \widetilde{\prod\limits_{\bold k}} \otimes
      \kett{0,\, \bold k, in}
  \nonumber\\
  &=&
      \widetilde{\prod_{\bold k}}  \otimes
      \left(  \inv{\vert \alpha_{k} \vert} \exp \left( z_{k}\,
      \hat  a_{\bold k}^{out \, \dagger}  \,
      \hat a_{-\bold k}^{out \, \dagger} \right)
      \ket{0,\, \bold k, out}\otimes \ket{0,\, -\bold k, out} \right) \, .
 \ea
The tilde tensorial product takes into account only half the
modes. It must be introduced because the squeezing operator acts
both on the $\bf k$ and the $-\bf k$ sectors. The definition of
this product requires the introduction of an arbitrary wave vector
to divide the modes into two sets. The sign of $k_z$ can be used.
Notice that a rigorous definition of $\widetilde{\prod_{\bold k}}$
requires to consider a discrete set of modes normalized with
Kroneckers (that is, to normalize the modes in a finite conformal
3-volume). To be explicit, the two-mode state $\kett{0,\, \bold k,
in} $ is given by
 \ba \label{in2mode}
   \kett{0,\, \bold k, in} =
   \ket{0,\, \bold k, in} \otimes \ket{0,\, -\bold k, in} \, ,
 \ea
where $\ket{0,\, \bold k, in}$ is the ground state of
 the $\bold k$-th mode at the onset of inflation. The complex parameter $z_k$
appearing in the squeezing operator in Eq. \Eqref{inoutvac} is
given by the ratio of the Bogoliubov coefficients
 \ba \label{z}
    z_k &=& \frac{\beta_k}{\alpha_k^*} = - e^{-2i k \eta_r}
    \frac{1}{\left( 1 + i2{k\eta_r}- 2k^2\eta_r^2 \right)}
    \nonumber \\
    &=& -  1 +  O(k\eta_r)^3  \, .
 \ea
 The high occupation number limit corresponds to
 $\abs{z_k} \to 1^-$.

It has to be emphasized that none of the out states in Eq.
\Eqref{inoutvac} carries 3-momentum. Hence, the distribution is
homogeneous in a strong sense: at late time the 3-momentum
operator is still annihilated by the state of Eq.
\Eqref{inoutvac}. (This property is not satisfied by incoherent
distributions such as thermal baths. In those cases, the
3-momentum fluctuates and vanishes only in the mean.) The present
distribution is also isotropic since the Bogoliubov coefficients
are functions of the norm $k$ only. Finally, it is a Gaussian
distribution, as can be seen from Eq. \Eqref{inoutvac}.

 To appreciate the peculiar properties of the distribution of Eq.
 \Eqref{inoutvac},
 and as a preparation for the analysis of the problem of
 decoherence and entropy,
 it is interesting to consider the most general homogeneous,
 isotropic, and Gaussian distributions of out quanta.
 A detailed description of these states is given in Appendix D.
 Their properties are completely specified by three real functions of
 the norm $k$ (one real and one complex) through the following
 expectation values
 \sba \label{momenta}
   && \langle \hat a_{\bold k}^{out} \rangle =  0 \, , \\
   &&\langle \hat a_{\bold k}^{out\, \dagger} \,
   \hat  a_{\bold k'}^{out} \rangle
   = n_k   \, \delta^3(\bold k - \bold k' ) \, ,   \\
   &&\langle \hat a_{\bold k}^{out}  \,
   \hat a_{\bold k'}^{out} \rangle
   = c_k \, \delta^3(\bold k + \bold k' )\, .
 \sea
In the second line, $n_k$ is the mean occupation number. In the
third one, the complex number $c_k$ characterizes the quantum
 coherence of the distribution.
The degree of two-mode coherence is given by $|c_k|/(n_k+1/2)$,
see Appendix C. It is bounded by 1. For a thermal (incoherent)
distribution, one has $c_k \equiv 0$: no 2-mode coherence.

 In the  case of pair production from vacuum, one has
 \ba \label{cterm}
   n_k = \abs{\beta_k}^2 = \frac{\abs{z_k}^2}{1- \abs{z_k}^2}\, ,
   \qquad
   c_k = \alpha_k \beta_k =
   \frac{z_k}{1- \abs{z_k}^2} \, .
 \ea
 Therefore, when considering relevant modes in inflation, using Eq.
 \Eqref{bogolincosmo}
 one has
  \sba \label{cterm2}
    n_k &=& \frac{1}{4 (k \eta_r)^4 } \simeq 10^{100}
    \, , \\
    c_k &=& - \left[ n_k + \frac{1}{2}+O(\frac{1}{n}) \right]
    e^{i2\theta_k} \, , \qquad
    \theta_k =  \frac{10}{3}(k\eta_r)^3  + O({n_k^{-5/4}}) \, .
 \sea
 That is, the distribution which results from inflation is highly
 populated \footnote{We recall how $n_k \simeq 10^{100}$ is obtained.
 For the test field $\hat \xi = \hat \phi/a$, using Eq.
 \Eqref{neglectdecay}, the power
 spectrum at the onset of the radiative era is, in physical units,
 \ba \label{defprimpower}
   {\cal P}_{\xi}(k; \eta_r) = \frac{k^3}{2\pi^2} \,
   \bra{0 in} \hat \xi_{\bf k}(\eta_r)
   \hat \xi_{\bf k}^{\dagger}(\eta_r) \ket{0 in} =
   \frac{k^3}{2\pi^2} n_k \times \frac{\hbar}{2k}
   \frac{4(k \eta_r)^2}{a_r^2} = \frac{k^3}{2\pi^2} \, \frac{\hbar}{2k}
    \, \frac{H^2}{k^2} \nonumber
    \, ,
 \ea
 where $a_r^{-1}= H \eta_r$.
 The power spectra of
 the gravitational potential $\Psi$ and the gravitons $h_{ij}^{TT}$,
 are obtained from
 the above. For slow-roll inflation \cite{MBF92},
 a mode by mode integration yields
 \ba \label{primpower}
   {\cal P}_{\Psi} = \frac{9}{25 \epsilon} \frac{4\pi \hbar G}{c^3}
   \left( \frac{H_k}{2\pi} \right)^2
   \left( 1 + O(\epsilon,\, \frac{d \ln \epsilon}{dN}) \right)
   \, , \qquad
   {\cal P}_{h_{ij}^{TT}} = 16 \epsilon {\cal P}_{\Psi} \, .
   \nonumber
 \ea
 Here $H_k$ designates the value of the Hubble parameter
 at horizon exit:
 $a(\eta_k) H(\eta_k) = k$.
 The data from COBE normalize the power spectrum
 to be $ \delta T_k / T \vert_{rms} \simeq \Psi_{k}/3
  \vert_{rms} \sim 10^{-5}$.
 The energy scale of inflation is thus
  $ (L_{Pl} H )^2 \simeq 10^{-10} \epsilon $,
 with $\epsilon = 10^{-2} - 10^{-1}$.
 The mean occupation number $n_k$ must therefore obey
 \ba
    n_{k} \simeq  \epsilon (\lambda_r H)^4 \, , \nonumber
 \ea
 where $\lambda_r = a_r/k$ is the physical wavelength of
 a mode $k$ at reheating. Assuming that reheating happened at
 redshift $\sim 10^{28}$,
 i.e. with a temperature $\simeq 10^{14} \rm{GeV}$,
 modes corresponding to large
 structures today, e.g. $100 {\rm Mpc} \sim 10^{21} {\rm m}$, have
 a mean occupation number $\simeq 10^{96}$.}
 and, more importantly, maximally coherent.
 These two properties go hand in hand for two-mode squeezed
 states.
 When computing the two-point function,
 the mean occupation number determines the
 {\it primordial}
 power spectrum, while
 the two-mode coherence of
the distribution manifests itself in the {\it temporal coherence}
of the modes, as we now explain.

\subsection{Two-point function and the neglect of the decaying mode}

 After the reheating,
 when expressed in terms of {\it out} modes, the (Weyl ordered) two-point
 function associated with the general distribution
 specified by Eqs. \Eqref{momenta} is
 \ba \label{Ginx}
    G^{in} (\eta,{\bf x}, \eta', {\bf x'}) &=&
     \int_0^{\infty} \frac{d k}{k} \frac{k^3}{2\pi^2}
    \frac{\sin(k\vert {\bf x} - {\bf x}'\vert)}{k\vert {\bf x} - {\bf x}'\vert}
    \left[ \left( n_k + \frac{1}{2} \right) \, \phi_k(\eta) \phi_k^*(\eta')
    +  c_k \, \phi_k(\eta) \phi_k(\eta') + c.c.
    \right]  \, . \nonumber \\
 \ea
We have dropped the superfix 'out' on the modes for lisibility. We
have chosen to work with the anti-commutator in order to obtain a
symmetrical function in $\eta,\, \eta'$ and to analyze the
classical limit, namely the high occupation number limit
 $n_k \gg 1$.

In order to discard the contribution of the decaying modes, two
conditions must be met.
 First, the sum in brackets should factorize.
 For a general distribution, it does not.
 However it factorizes
 for coherent states, see Eq. \Eqref{CohWaveFunct} in Appendix A,
 for distributions resulting from pair production from vacuum
 in the high occupation number limit,
 and, more generally, whenever the following inequality is satisfied
 \ba \label{neqqq}
   \frac{n_k+1/2 - \abs{c_k} }{n_k + 1/2} \ll 1 \, .
 \ea
 Let us write
 \ba \label{epsilontheta}
  c_k = - e^{i2\theta_k} (n_k+ \frac{1}{2})\, (1-\epsilon_k)
  \, ,
 \ea
 where $\epsilon_k$ and $\theta_k$ are real.
 Eq. \Eqref{neqqq} is equivalent to  $\abs{\epsilon_k} \ll 1$. Then
 one can write the 
 bracket of \Eqref{Ginx} as
 \ba \label{brackets}
   \left( n_k + \frac{1}{2} \right)\left[
   \left( e^{i\theta_k} \phi_k(\eta) - c.c. \right) \,
   \left( e^{-i\theta_k} \phi^{*}_k(\eta') -  c.c. \right) \, + \,
   \epsilon_k  \left( e^{i2\theta_k} \phi_k(\eta) \phi_k(\eta') +
   c.c. \right)
   \right] \, .
   \label{powern}
 \ea
One thus verifies that Eq. \Eqref{neqqq} is sufficient to
factorize the two-point function.

The second condition for discarding the decaying mode
 arises from the fact that the mode in parenthesis,
 \ba \label{nonalwaysgrowingmode}
    \phi_k(\eta) = i2Im\left( e^{i\theta_k} \phi^{out}_k(\eta) \right) \, ,
 \ea
will not, in general, be the growing mode. Indeed
it is only for $\theta_k \ll 1$ that one
can approximate it by the sine function.

In brief, to discard the cosines, both $\epsilon_k \ll 1$ and
$\theta_k \ll 1$ must be satisfied. When considering pair creation
from vacuum \Eqref{bogolincosmo}, we get
 $\epsilon_k \propto 1/n_k$ and
 $\theta_k \propto 1/n_k^{3/4}$. Hence it is perfectly legitimate
 to discard the cosines.
In that case, the two-point
function of the physical field $\xi = \phi /a$ reduces to
 \ba \label{Gxi}
   \bra{0 in} \{ \, \hat \xi(\eta, {\bf x}),\,
   \hat \xi(\eta', {\bf x}') \,\}\ket{0 in}
   &=&
   \int_0^{\infty}\!\! \frac{d k}{k} \,
   \frac{\sin(k\vert {\bf x} - {\bf x}'\vert)}{
          k\vert {\bf x} - {\bf x}'\vert} \,
          {\cal P}^{0}_{\xi}(k) \,
          \frac{\sin (k\eta )}{k \eta}
          \frac{\sin (k \eta')}{k\eta'}
   \, . 
 \ea

  The remaining statistical properties of the distribution are the
 power spectrum and the temporal coherence of the growing mode.
 First, the primordial power spectrum
 \ba \label{primpowerxi}
    {\cal P}^{0}_{\xi}(k) = \left( \frac{H}{2\pi} \right)^2 \, ,
 \ea
 is proportional to
 the mean occupation number $n_k$,
 see footnote 2.
 Here, it is scale
 invariant because the inflationary background has been approximated by
 de Sitter space.
 Second, the time dependent function which appears in \Eqref{Gxi} is
 $(\sin k\eta)/(k\eta)$,
 where $\eta$ is proportional to the
 scale factor given by Eq. (\ref{aeta}b).
 This is how the temporal coherence of modes obtains
 from the two-mode coherence of the distribution. Consequently,
 the power spectrum ${\cal P}_{\xi}(k; \eta)$ at time
 $\eta$, defined by the Fourier transform of the two-point
 function,
 is the product of the primordial power spectrum and the
 square of the mode function evaluated at that time:
 \ba \label{power}
   {\cal P}_{\xi}(k; \eta) &=& {\cal P}^{0}_{\xi}(k) \,
   \left( \frac{\sin k \eta}{k\eta} \right)^2 \, .
 \ea
The power
 spectrum at a given time, as a function of k, oscillates and has
 zeros. This behaviour
 is a necessary condition for
 the existence of acoustic oscillations and anti-correlations in
 the temperature anisotropy TT and cross-correlation TE power
 spectra respectively \cite{WMAP1}.

 Hence, once the cosine is neglected, the quantum distribution
 can be
 effectively replaced by a stochastic Gaussian distribution of
 classical fluctuations
 \ba \label{sineeq}
   \xi_{\bf k}(\eta) = \xi^0_{\bf k} \, \frac{\sin k\eta}{k\eta}  \, ,
 \ea
 with locked temporal argument, and random amplitudes with
 variances given by
 \ba \label{Aeff}
   \av{\xi^0_{\bf k} \, \xi_{\bf k'}^{0\, *}}_{e\!f\!f} =
   \av{\xi^0_{\bf k}\, \xi^0_{-\bf k'}}_{e\!f\!f} =
   {\cal P}_{\xi}^{0}(k) \, \delta^{3}({\bf k}-{\bf k}') \, .
 \ea
 Being Gaussian, the effective probability distribution is simply
 \ba \label{Peff}
  {\cal P}_{e\!f\!f} = \widetilde{\prod_{\bold k}} \,
  \frac{1}{{\cal P}_{\xi}^{0}(k)}
  \exp\left(-\frac{\abs{\xi^0_{\bf k}}^2}{{\cal P}_{\xi}^{0}(k)} \right) \, .
 \ea
 To avoid double counting, one must again
 use the tilde product which takes into account half the modes
 only, as was done in the quantum distribution of Eq. \Eqref{inoutvac}.
 This counting becomes crucial when computing the entropy, see Section IV.C.

 We emphasize that the reality of the field $\phi(\eta,{\bf x})$
 has nothing to do with the temporal coherence of the 
 modes of Eq. \Eqref{nonalwaysgrowingmode}.
 Nevertheless, once having neglected the decaying mode, it is true that
 the first equality
 in Eq. \Eqref{Aeff}
 is imposed by the reality of $\phi(\eta,{\bf x})$.
 However, this should not be confused with the relation between
 the expectation values of the creation and destruction operators
 in the squeezed vacuum state prior to have discarded the
 decaying mode,
 \ba \label{newn}
   \langle \hat a_{\bf k}^{out\, \dagger}
   \, \hat a_{\bf k'}^{out} \rangle = -
   \langle \hat a_{\bf k}^{out}  \, \hat a_{-\bf k'}^{out} \rangle
   (1 + O(k \eta_r)^3)\, .
 \ea
 The latter follows from Eqs. \Eqref{momenta}-\Eqref{cterm2}, and
 is the expression of the {\it two-mode coherence} of the
 {\it in} vacuum (or more generally of strongly correlated
 two-mode distributions, see Section V).
The (complex) Eq. \Eqref{newn} guarantees
 that both conditions $\epsilon_k \ll 1, \, \theta_k \ll 1$ are
 satisfied, thereby allowing
 to factorize the 2-point function
 and to discard the contribution of the
 decaying modes.

\subsection{Additional remarks}

First we remind the reader why a random distribution of both the
sine and the cosine does not give rise to temporal coherence
 \cite{Grishchuk90}. In fact such a distribution  corresponds to an
incoherent (thermal) distribution.

Consider the incoherent distribution:
 \ba \label{incoh}
   \langle \hat a_{\bf k}^{\dagger} \hat a_{{\bf k}'}
   \rangle_{i\!n\!c} &=& n_k \, \delta^{3}({\bf k}-{\bf k}') \, ,
   \nonumber \\
   \langle \hat a_{\bf k} \hat a_{{\bf k}'} \rangle_{i\!n\!c} &=&
   c_k \, \delta^{3}({\bf k}+{\bf k}') = 0 \, .
 \ea
The last equation implies that temporal coherence of the modes is
lost, as can be seen from the temporal behaviour of the bracket in
the integrand of Eq. \Eqref{Ginx} : when $\eta = \eta'$  the
bracket no longer exhibits oscillations in $k$.
 (This absence can also be understood by
 considering
 field amplitudes as classical stochastic variables
 rather than quantum ones.
 Writing the mode in terms of its norm and its phase
  \cite{Grishchuk90,Allen00}
 \ba
   \phi_{\bf k} &=& \Phi_{\bf k} \, \sin(k\eta + \theta_{\bf k})
   \, ,
 \ea
 one can treat $\Phi_{\bf k}$ and $\theta_{\bf k}$ as stochastic
 variables. One verifies that the distribution for the phase
 $\theta_{\bf k}$ is uniform over the interval
 $\left[0,\, 2\pi \right]$.
 Hence, taking the ensemble average to compute the power spectrum,
 no temporal coherence could obtain.)

 It is of value to estimate what has been neglected when
 discarding the cosine in Eq. \Eqref{sineeq}.
  To this end, we
 write the field modes
 directly in terms of the growing and decaying modes
 \ba \label{gdmodes}
   \hat \phi_{\bf k} &=& \hat a_{\bf k}^{out} \phi_k^{out} +
   \hat a_{-\bf k}^{\dagger \, out} \phi_k^{out \, *} \, \nonumber \\
   &=& \hat g_{\bf k} \frac{\sin(k\eta)}{\sqrt{k}}
   +\hat d_{\bf k} \frac{\cos(k\eta)}{\sqrt{k}}\,  .
 \ea
 The leading non-vanishing order in $n_k$ of their fluctuations
 and cross correlation are
 \sba \label{ABvariances}
   \langle \hat g_{\bf k} \hat g_{{\bf k}'}^{\dagger} \rangle_{in} &=&
   \left( n_k + \frac{1}{2} - Re(c_k) \right) \, \delta^{3}({\bf k} - {\bf k}')
   = 2 n_k  \,
    \left( 1 + O(n_k^{-1})  \right) \, \delta^{3}({\bf k} - {\bf k}')
   \, , \\
   \langle \hat d_{\bf k} \hat d_{{\bf k}'}^{\dagger} \rangle_{in} &=&
   \left( n_k + \frac{1}{2} + Re(c_k) \right) \, \delta^{3}({\bf k} - {\bf k}')
   =  O(n_k^{-1/2})
    \, \delta^{3}({\bf k} - {\bf k}')
   \, , \\
   \langle \{ \hat d_{\bf k}, \hat g_{{\bf k}'}^\dagger \} \rangle_{in} &=&
   Im(c_k) \, \delta^{3}({\bf k} - {\bf k}')
   = O(n_k^{1/4})
   \, \delta^{3}({\bf k} - {\bf k}')
   \, .
 \sea
 Notice that the cross-correlation,
 $\langle \{ \hat d_{\bf k}, \hat g_{{\bf k}'}^\dagger \} \rangle_{in}$,
 the Weyl ordered product (i.e. the anticommutator
 divided by 2),
 is non-zero. This means
 that the growing and the decaying mode
 are not eigenmodes of the distribution (see Appendix C).


 When divided by the variance of $\hat g_{\bf k}$, the variance
 of 
 the cross-correlation is of order
 $n_k^{-3/4} \sim (k\eta_r)^3 \sim 10^{-75}$. 
 The power of the decaying mode is even smaller.
 One can therefore
 safely use the effective distribution Eq. \Eqref{Peff} in replacement of the
 original quantum distribution Eqs. \Eqref{inoutvac} or \Eqref{ABvariances}
 when calculating the power spectrum. However this is {\it not} the
 case for the entropy.

 It is illustrative to compare the in-vacuum
 correlations, Eqs. \Eqref{ABvariances},
 to those
 of the incoherent distribution \Eqref{incoh}.
 In that case
 $\hat d_{\bf k}$ and $\hat g_{\bf k}$ are uncorrelated Gaussian
 operators with equal variance:
 \ba \label{incoherent}
    \langle  {\hat g_{\bf k}\hat g_{{\bf k}'}^\dagger}\rangle_{i\!n\!c}&=&
    \langle {\hat  d_{\bf k}\hat d_{{\bf k}'}^\dagger}\rangle_{i\!n\!c}
   =
   \left( n_k + \frac{1}{2} \right) \, \delta^{3}({\bf k}-{\bf k}')
   \, , \nonumber \\
   \langle
   \{\hat d_{\bf k} , \hat g_{{\bf k}'} \} \rangle_{i\!n\!c} &=& 0
   \, .
 \ea
These differences with Eqs. \Eqref{ABvariances}
are particularly clear.

\subsection{Drawbacks of the simplified description}

The simplified description in terms of a statistical ensemble of
sine standing waves, see Eq. \Eqref{Peff}, has several
shortcomings. It is of value to describe them with some attention

First, there is the discrepancy of the value of the entropy
mentioned after Eq. \Eqref{ABvariances}. Second, it should be
pointed out that, in the early universe, non-linear processes,
however weak they may be, will modify the density matrix obtained
by making use of free fields. To be able to parameterize the
modifications of the power spectrum, one must return to
 two-mode distributions \footnote{As mentioned in the Introduction,
 we shall neglect
 non-Gaussianities when computing the entropy.
 In other words, we replace the actual (non-Gaussian) distribution
 by the Gaussian one which possesses the same
 moments \Eqref{momenta}
 and therefore the same power spectrum.
 The justifications for this substitution are the following.
 First, it leads to a drastic simplification of the analysis, and
 {\it all} Gaussian (isotropic) distributions can be considered.
 Second, when focussing on a given wave vector, one anyway traces
 over the
 non-Gaussianities relating this scale to other wave vectors.
 Third, since non-Gaussianities are expected to be
 weak \cite{Maldacena03,Matarrese}, it is a well defined
 (mean field/Hartree) and useful
 approximation to replace the actual distribution
 by the Gaussian one associated to it.}.
 Indeed, the classical ensemble of sine waves
 cannot describe these modifications. It should be considered only
 as an effective description of the density matrix when the
 decaying
 mode is strictly negligible.

The appropriate basis to describe density matrices is provided by
coherent states. The reasons are the following. First, they
constitute the quantum counterparts of classical configurations in
phase space. Therefore they provide an adequate basis for studying
the semi-classical limit. In particular, when non-diagonal terms
are still present in this basis, it is the signal that some
quantum coherence has been kept,
 see Section V and Appendix D.3.
 Secondly,
 they are the preferred basis in which squeezed states
decohere when weak interactions are taken into account,
 see Section IV.A and \cite{Zurek93} for further details.



\section{Two-mode coherent states}

 When using coherent states in inflationary cosmology, one must pay attention to
 the entanglement between $\bf k$ and $-{\bf k}$ modes.
 As we shall see this leads to the notion of
 'two-mode coherent states'.
 On the contrary, a naive use of coherent states assigning
 amplitudes to each mode separately
 will erase the entanglement and therefore suppress the
 temporal coherence of the modes.
 For completeness this is shown in the next subsection.

 \subsection{Naive use of coherent states and the loss of
 temporal phase}

 If assuming that each mode decoheres separately, the matrix
 density of the ${\bf k}, {\bf -k}$ pair of modes factorizes
 and the modes are independent:
 \ba \label{naive}
   \hat \rho^{Naive}
   = \left( \int\!\!\frac{d^2 v}{\pi} \, G(v)
   \ket{v, \, {\bf k}} \bra{v, \, {\bf k}} \right)
   \otimes
   \left( \int\!\!\frac{d^2 w}{\pi} \,
    G(w) \ket{w,\, - {\bf k}} \bra{w,\, - {\bf k}} \right) \, ,
 \ea
 where $G(v) = n_k^{-1} \exp\left( - \abs{v}^2/n_k \right)$
 is the normalized Gaussian distribution giving the
 appropriate power-spectrum,
 and where $\ket{v, \, {\bf k}}$ and
 $\ket{w,\, - {\bf k}}$ are coherent states
 constructed with out operators, see Appendix A for their definition.
 The expectation values in this decohered density matrix are:
 \ba \label{RLcoherence}
   {\rm Tr}(\hat \rho^{Naive}  \hat a_{\bf k}^{\dagger} \hat a_{\bf k} ) &=&
   n_k   \, ,  \nonumber\\
   {\rm Tr}(\hat \rho^{Naive}  \hat a_{\bf k} \hat a_{-{\bf k}} ) &=& 0\,  .
 \ea

 One thus verifies that Eq. \Eqref{naive}
 corresponds to the incoherent distribution of Eq. \Eqref{incoh}
 written in the basis of one-mode coherent states.
 The two-mode correlations are completely
 erased and the temporal coherence of the modes is lost.
 In fact Eq. \Eqref{naive} is maximally decohered,
 since it is a tensorial product of two thermal states
 (because $G(v)$ is Gaussian and isotropic in the complex $v$ space).

 \subsection{Representation of the in vacuum with two-mode coherent states}

 To understand the usefulness
 of two-mode coherent states it is appropriate to first mention
the following properties \cite{CP04}. Consider a mode ${\bf k}$ in
a coherent state $\ket{v,\, {\bf k}}$. Then compute the one-mode
reduced state obtained by projecting it on the two-mode initial
vacuum of Eq. \Eqref{in2mode}. You will get:
 \ba \label{remar}
   \langle v,\, {\bf k} \kett{0 in,\, {\bf k}}  =
   {\cal A}_k(v) \, \ket{z_k v^*, \, -{\bf k} } \, .
 \ea
It is remarkable that the state of the $-{\bf k}$ mode which is
entangled to $\ket{v,\, {\bf k}}$ is also a coherent state. Its
amplitude is given by the complex conjugate of $v$ times $z_k$
characterizing the pair creation process. These facts follow from
the EPR correlations in the initial vacuum displayed in Eq.
\Eqref{inoutvac}. The prefactor ${\cal{A}}_k(v)$ is
 \ba \label{A(v)}
   {\cal{A}}_k(v) &=& \inv{\abs{\alpha_k}}
   \exp{\left(-\frac{\abs{v}^2}{2\abs{\alpha_k}^2} \right)}  \, .
 \ea
It is the probability amplitude to find the mode ${\bf k}$ in
the coherent state $\ket{v, \, {\bf k}}$ given that we start with vacuum at the
onset of inflation.

Using the representation of the identity with coherent states, Eq.
\Eqref{coh1}, Eq. \Eqref{remar} permits to decompose the two-mode
{\it in} vacuum as a {\it single} sum of {\it two-mode coherent
states}. It should be stressed that two independent integrations
over one-mode coherent states are necessary to describe a generic
two-mode state. The fact that only one integration is sufficient
is therefore a direct consequence of the two-mode coherence of the
initial state $\kett{0 in,\, {\bf k}}$. Explicitly we have
 \ba \label{insuperpcoh}
   \kett{0 in,\, {\bf k}}  =
   \int\!\!\frac{d^2v}{\pi} \,
   {\cal{A}}_k(v) \, \,  \ket{v,\, {\bf k}}  \otimes \ket{z_k v^*, \, -{\bf k}}
   \, .
 \ea
 This result is exact
 \footnote{Notice however that the above
 decomposition is not unique since the coherent states are not
 orthogonal, compare with \cite{Kim00}.
 Eq. \Eqref{insuperpcoh} has the advantage to be directly related to
 the detection
 of a quasi-classical configuration in the ${\bf k}$ sector.}
 and applies even for low occupation numbers.
 It is the consequence of the coherence of
 the in vacuum and
 holds for every homogeneous pair creation process.

 Another important consequence of Eq. \Eqref{remar} is that the
 probability to find {\it simultaneously} the ${\bf k}$-mode with
 coherent amplitude $v$ and its partner with amplitude $w$ is
 \ba \label{vwdistrib}
   {\cal P}_{2, \, k}(v,w) =
   \abs{\bra{v,\, {\bf k}}\bra{w,\, -{\bf k}} 0 in, \, {\bf k} \rangle_2}^2 =
   \abs{{\cal{A}}_k(v)}^2 \times e^{-\abs{w-z_k v^{*}}^2} \, .
 \ea
 The second factor follows from the overlap between the reduced
 state in the r.h.s. Eq. \Eqref{remar} and the bra
 $\bra{w,\, -{\bf k}}$
 (we also used the coherent state overlap:
 $\abs{\langle u \vert v \rangle}^2 = \exp({-\abs{u-v}^2}) $).

Equation \Eqref{vwdistrib} implies that once the amplitude of the
${\bf k}$-mode is fixed, the conditional probability to find its
partner in a coherent state $\ket{w}$ is centered around
 \ba \label{meanw}
  \bar w(v)=z_k v^{*} \, .
 \ea
Moreover, in the high occupation number limit we are dealing with,
the spread ($=1$) around this mean value is negligible when
compared to the spread in $v$ which is given by $\abs{\alpha_k}^2
= n_k + 1$. Therefore, when computing expectation values to
leading order in $n_k$, the conditional probability acts as
 $\delta^{(2)}(w- \bar w(v))$.
Hence both the real and the imaginary parts of $w$ are fixed. This
is how the EPR correlations in the in-vacuum translate in the
coherent states basis: double integrations over one-mode coherent
states can be replaced by single integrations over two-mode
states.

 Furthermore, the strict correspondence between the coherent
 state amplitudes in the sectors ${\bf k}$ and $-{\bf k}$
 leads, in the high-squeezing limit,
 to the temporal coherence of modes since the only random variable
 is their common amplitude
 \cite{CP04}.
 As we shall see below, this result is
 crucial as it allows to erase the quantum
 features of the distribution (such as the existence of a
 sub-fluctuant direction in phase space)
 while preserving the temporal coherence
 of modes upon horizon re-entry (which can be classically interpreted).

 To complete this analysis, and in preparation for studying decoherence,
 it is also interesting to explicitly write the non-diagonal matrix
 elements of the in vacuum density matrix. One has
 \ba \label{vwrhov'w'}
   \bra{v}\bra{w} \hat \rho_{in} \ket{v'} \ket{w'} &=&
   {\cal A}_{2, \, k}(v,w) \, {\cal A}_{2, \, k}(v',w')^{*} \, ,
 \ea
 where the two-mode amplitude is
 \ba
   {\cal A}_{2, \, k}(v,w) &=& {\cal A}_k(v) \,
   e^{-\frac{1}{2}\abs{w-z_k v^*}^2 }\,
   e^{i\, {\rm Im}(w^*z_k v^*)} \, .
 \ea
Since the initial vacuum is a pure state and since
$\abs{\alpha_k}^2 = n_k + 1 \gg 1$, the above matrix elements do not vanish
when $w = z_k v^*, \, w' = z_k {v'}^*$ even for $\abs{v-v'}^2 \simeq n_k$,
i.e. they do not vanish
{\it even for macroscopically different coherent states}.
(Remember that $n_k \simeq 10^{100}$.)

Because of this macroscopic quantum coherence, $\hat \rho_{in}$
does not describe a classical ensemble (a mixture) of coherent
states. Fortunately, as we shall see below, this coherence is
unstable to small modifications of the distribution. To describe
this decoherence we shall proceed in two steps. First we shall
consider a specific example of decohered density matrix and argue
why it plays a critical role, namely, it separates distributions
which have kept some quantum coherence from those which are
mixtures (i.e. distributions for which the off-diagonal matrix
elements vanish for $\abs{v-v'}^2 \geq 1$.) Second, we shall
describe partially decohered density matrices in general and more
technical terms. In particular we shall sort out the robust
properties from its unstable ones. In one word, when using
$|c_k|/(n_k + 1/2)$ as a parameter for characterizing the two-mode
coherence of the distribution, we shall see that the power
spectrum and the temporal coherence are robust whereas the
entropy, the above quantum coherence, and the residual squeezing
are extremely sensitive to this parameter. Even though the value
of this parameter will be introduced from the outset, similar
results obtain when performing a dynamical analysis.

 \section{Minimal decoherence scheme of primordial spectra}

\subsection{Zurek {\it et al.} analysis and minimal decoherence scheme}

In general, it is a difficult question to determine into what
mixture an initial density matrix will evolve when
 taking into account some interactions with other modes
 and tracing over these extra degrees of freedom so as
 to obtain an effective (reduced) density matrix.
There exist  however several cases where clear conclusions can be
drawn. First, when one can neglect the free Hamiltonian, the
preferred states (that is the basis into which the reduced density
matrix will become diagonal) are the eigenstates of the
interaction Hamiltonian \cite{gottfried,Zurek81,Zurek82}. This
approach has been applied in \cite{KiefPolStar98a}, to primordial
density fluctuations when the (physical) modes are almost constant
because their wave length is much larger than the Hubble radius.
The conclusion is that the preferred basis is provided by
amplitude (position) eigenstates. However this conclusion leaves
some ambiguity and might lead to some difficulties. First,
position eigenstates are not normalized. Second, and more
importantly, the spread in momentum is infinite for these states.
Therefore, the velocity field would not be well defined when the
modes re-enter the horizon. Moreover, as pointed out in
 \cite{KiefPolStar00}, some additional decoherence could be
obtained as they re-enter the horizon. In this case, the momentum
should be treated in the same footing as the position. To cure
these problems,
 some finite spread in position should be introduced.
One then needs a physical criterion to choose this spread.

To introduce a critical spread,
 to identify the various regimes of decoherence,
 and to determine their range, one should use
coherent states, as we now
argue. (We reserve for a forthcoming publication a proper
justification of the physical relevance of these states in a
cosmological context.)
 First, as shown in \cite{Zurek93}, when considering
harmonic oscillators weakly interacting with an environment,
coherent states provide the basis in which the density matrix
decoheres. In particular,
 when the initial state is
 a squeezed state, there is a phase of rapid growth of the entropy
 which sends the system into a mixture of coherent states.
 This is then followed by a period of slower increase accompanying
 dissipation or thermalization.
 Second, in inflationary cosmology, fluctuation modes
 are weakly interacting harmonic oscillators \cite{Maldacena03}.
 Indeed, given that the relative density
 fluctuations have small amplitude ($\sim 10^{-5}$), the hypothesis
 of weak interactions is perfectly legitimate.
 Therefore, the modes with ${\bf k'} \neq {\bf k} $ will effectively
 act as an environment for a given two-mode state.
A novelty with respect to the analysis of \cite{Zurek93} is that
in cosmology, one deals with two-mode squeezed states. In this
case one ends up with statistical mixtures of two-mode coherent
states, but not with mixtures of one-mode coherent states. The
procedure to define these distributions
is explained in Appendix B.
However, because coherent states are
overcomplete, there is an inherent ambiguity (albeit small in the large $n$
limit) in determining this distribution,
{\it c.f.} footnote 4 for a related 
ambiguity. The procedure
we shall adopt in this Section is to use Eq. \Eqref{vwdistrib} to
define it.


 So unless the interactions taking place in the early cosmology
 are sufficiently weak (or sufficiently anisotropic
 in phase space
 as in the case they act only for a fraction of a period)
 so as to keep some quantum coherence,
 we can infer that the
 effective entropy (see footnote 3)
 of the final distribution should be higher than (or equal
 to) that stored in
 the diagonal distribution of two-mode coherent states,
 here after called minimally decohered.

 In the next two subsections, we shall explicitly write down this
 distribution, compute the entropy it carries, and, very
 importantly, show that the temporal coherence of the modes is kept
 whereas the macroscopic quantum coherence of the
 original distribution, see Eq. \Eqref{vwrhov'w'}, is lost, thereby
 allowing a classical interpretation of the residual statistical
 properties.

\subsection{The minimally decohered distribution}

The minimally decohered distribution which corresponds to Eq.
\Eqref{vwrhov'w'} is
 \ba \label{rhodec}
   \hat \rho_{min} = \int\!\!\frac{d^2v}{\pi} \frac{d^2w}{\pi}\,
   {\cal{P}}_{2, \, k}(v,w) \, \,
   \ket{v,\, {\bf k}}\bra{v,\, {\bf k}} \otimes
   \ket{w, \, -{\bf k} }\bra{w, \, -{\bf k}} \, .
 \ea
The probability distribution  ${\cal{P}}_{2, \, k}(v,w)$
is given in Eq. \Eqref{vwdistrib}.

By having replaced the original distribution
$\hat \rho_{in}= \ket{0, in}\bra{0, in}$ by the decohered one
$\hat \rho_{min}$, some entropy has been introduced.
Indeed, by direct computation, one gets
 \ba \label{RLcoherence}
   {\rm Tr}(\hat \rho_{min} \hat a_{\bf k}^{\dagger} \hat a_{\bf k} ) &=&
   \abs{\alpha_k}^2 = n_k + 1
   \, ,
   \nonumber\\
   {\rm Tr}(\hat \rho_{min} \hat a_{\bf k} \hat a_{-{\bf k}} ) &=&
   z_k \abs{\alpha_k}^2 = c_k \,  .
 \ea
The first line shows the occupation number increased by one unit.
Thus the relative change is only $1/n_k$. In the second line one
finds that that the
 coherence term
 is equal {\it in phase and amplitude} to that of the
original distribution, see Eqs. (\ref{momenta}, \ref{cterm}).
These two results imply, first, that the (relative) degree of
coherence has been reduced and therefore some entropy has indeed
been gained. Secondly, in the high occupation number limit we are
dealing with, the two-point function of Eq. \Eqref{Ginx} is not
affected by this loss of coherence since, for relevant modes,
 the relative change is of the order of $1/n_k \sim 10^{-100}$.
 Let us emphasize that the smallness of this change
 follows from the fact that the phase of
 ${\rm Tr}(\hat \rho_{min} \hat a_{\bf k} \hat a_{-{\bf k}} )$
 stays equal to that of the r.h.s of Eq. (\ref{momenta}c).
 Hence, the fact that Eq. \Eqref{neqqq} still applies leads
 to the preservation of the temporal coherence of modes.

 In fact, the main modification associated with the replacement
 of
 $\hat \rho_{in}= \ket{0, in}\bra{0, in}$ by
 $\hat \rho_{min}$ concerns the elimination of the
 non-vanishing off-diagonal matrix elements.
 Indeed, in the high
 squeezing limit, we have, see also Appendix D.$2$,
 \ba
   &&\bra{v}\bra{w} \hat \rho_{min} \ket{v'}\ket{w'}
   \simeq \frac{1}{3 \abs{n_k}} \times \nonumber \\
   &&\exp-\frac{1}{2} \left[ \abs{v}^2 + \abs{v'}^2 + \abs{w}^2 +
   \abs{w'}^2 - \frac{4}{3}\left(v^* v' + w^* w' -
   \frac{z_k}{2}v^* w^* - \frac{z_k^*}{2}v w \right) \right]
   \, .
 \ea
 (We have set $\abs{z_k} = 1$ at the end
 of calculations to simplify the expression.)
 The above result implies that along the
 big axis of the ellipse, see Fig. 1,
 i.e. for $w=z_kv^*$, the term in the exponential is $ (4/3) \abs{v-v'}^2$.
 Therefore,
 two coherent states separated by $\abs{v' - v} > 1$ will no
 longer interfere quantum mechanically.

 Moreover, each one of the coherent states in Eq. \Eqref{rhodec} is
 stable under the evolution of the quadratic Hamiltonian, i.e., the
 expectation values of
 $\hat \phi_{\bf k}, \, \partial_{\eta} \hat \phi_{\bf k}$ in these states
 evolve according to the classical equations of motion, see
 Eqs. \Eqref{cltrajincohstate}.
 Hence, the
 diagonal distribution \Eqref{rhodec} can safely
 be interpreted in classical statistical mechanics, i.e. as a
 (stable) statistical ensemble of (non-interacting) classical states.

Finally, when computing expectation values in leading order in
$n_k$, this distribution can be further simplified by replacing
the Gaussian factor of unit spread of Eq. \Eqref{vwdistrib} by a
double delta function
 as discussed
 after Eq. \Eqref{meanw}.
 Using this equation,
 one gets the simplified distribution
 \ba \label{rhodec2}
   \hat \rho_{simpl} = \int\!\!\frac{d^2v}{\pi} \,
   \abs{{\cal A}_k(v)}^2  \,\,
   \ket{v,\, {\bf k}} \bra{v,\, {\bf k}}
   \otimes
   \ket{\bar w(v), \, -{\bf k}} \bra{\bar w(v), \, -{\bf k}}
   \, .
 \ea
It is now written as a single sum of two-mode coherent states,
thereby making explicit the fact that we are dealing with a
statistical superposition of two-mode states (and not a double
superposition of one-mode coherent states).

 \begin{figure}[ht] \label{decoherence}
 \epsfxsize=15.0truecm
 \epsfysize=12.0truecm
 \centerline{{\epsfbox{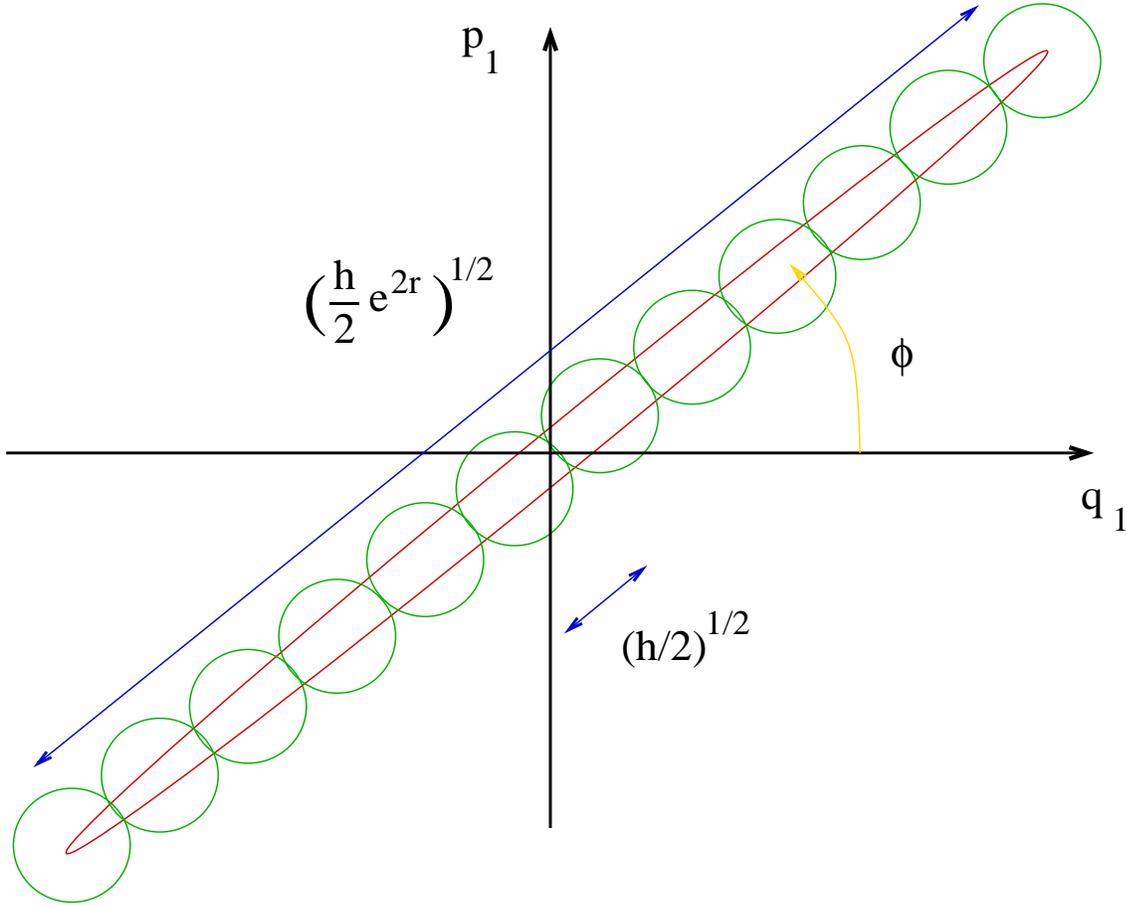}}}
 \caption{{\it Decoherence  of a one-mode squeezed state in the coherent
 state basis.}
 Given the 2-dimensional character of this sheet of paper, it is
 simpler to represent the phase space of a one-mode squeezed state
  than that of a two-mode squeezed state. To this end, one should
 first decompose the two-mode squeezed states we are dealing with into two
 copies of one-mode squeezed states, as exposed in Appendices B and D.4.
 The large red ellipse represents the
 $1-\sigma$ contour of a one-mode squeezed state in $(q_1,\, p_1)$ space,
 with $\omega = 1$.
 The squeezed state is characterized by the squeezing parameters
 $r_k$ et $\phi_k$, where $n_k = \sh^2 r_k$ and $\phi_k = \arg(c_k)/2$.
 The small green circles represent the $1-\sigma$
 contour of coherent states, which have no privileged direction
 in phase-space. When replacing $\hat \rho_{in}$ by $\hat \rho_{min}$,
 two modifications obtain. First, the squeezed spread in the direction of the
 small axis of the ellipse is replaced by that of the vacuum. Second,
 and concomitantly, the macroscopic quantum
 coherence along the big axis, see Eq. \Eqref{vwrhov'w'},
 is removed and replaced by that of coherent states.}
 \end{figure}

\subsection{Minimal entropy and the neglect of the decaying mode}

Our aim is to compute the entropy stored in Eq. \Eqref{rhodec}.
The entropy of any Gaussian two-mode distribution,
$\hat \rho_{{\bf k},\, -{\bf k}}$,  can be exactly calculated
\cite{Zeh85,Serafini04} by using the fact that its
density matrix is unitarily 
equivalent to the tensorial product of
two thermal density matrices of auxilliary oscillators, see Appendix C.
 These two real oscillators can be taken to be the real
 and the imaginary parts of $\phi_{\bf k}$ we used in Appendix B.
 One can indeed write
 \ba \label{map}
   \hat \rho_{{\bf k},\, -{\bf k}} = {\cal M}^{\dagger} \, \hat \rho_{th,\, 1} \otimes
   \hat \rho_{th,\, 2} \, {\cal M} \, ,
 \ea
 where $\cal M$ is a unitary operator acting on the two-mode Hilbert space.
 The expression of the (von Neumann) entropy immediately follows:
 \ba
   S\left[ \hat \rho_{{\bf k},\, -{\bf k}} \right] &=& S\left[ \hat \rho_{th,\, 1} \right] +
   S\left[ \hat \rho_{th,\, 2} \right] \, ,
 \ea
 where the entropy of a thermal bath with mean occupation $\bar n$ is
 \ba  \label{Stherma}
   S\left[ \hat \rho_{th} \right] = (\bar n +1 ) \ln(\bar n +1 ) -
   \bar n \ln(\bar n) \, .
 \ea
When considering 
distributions preserving homogeneity and
isotropy, the occupation numbers of the thermal matrices are equal
and given by
 \ba \label{tilde n}
   \bar n_{k} + \frac{1}{2} =
   \left( (n_{k} + \frac{1}{2})^2 - \abs{c_{k}}^2 \right)^{1/2} \, ,
 \ea
where $n_k$ and $c_k$ are defined in Eq. \Eqref{momenta}. It
should be noticed that the phase of $c_k$ (which is essential for
the temporal coherence of modes) drops out from this expression.
Hence the quantum purity is not univocally related to the temporal
coherence.

Let us apply Eqs. (\ref{Stherma}, \ref{tilde n}) to several cases.
First, for the two-mode in vacuum of Eq. \Eqref{inoutvac}, the
occupation number and the coherence term are related by Eq.
\Eqref{cterm}, one has
 $\bar n_k = 0$, as expected. Hence the entropy vanishes.

For the decohered matrix Eq. \Eqref{rhodec}, using Eq.
(\ref{RLcoherence}), the occupation number of the two thermal
baths are
 \ba \label{barn}
   \bar n_{k} =
   \frac{1}{2} \left( -1+\sqrt{8(n_k +1 ) +1}\right) \sim
   \sqrt{2  n_k} \, ,
 \ea
where the last term is the leading order when $n_k \gg 1$. The
two-mode entropy of this mixture is then
 \ba \label{Smin}
   S_{{\bf k},\, -{\bf k}}\left[ \hat \rho_{min} \right] &=&
   2 \, S\left[ \hat \rho_{th}
   \right] = 2 \ln \bar n_k + O(1) \, ,
   \nonumber \\
   &=& \ln n_k + O(1) = 2\, r_k +  O(1) \, ,
   \nonumber \\
   &\simeq& 100 \ln(10)  \, .
 \ea
  In the second line, we have expressed the occupation number in
term of the squeezing parameter $r_k$ : $n_k = \sh^2 r_k$. Hence,
a two-mode squeezed vacuum state which decoheres in the two-mode
coherent basis goes along with an entropy of $S_{{\bf k},\, -{\bf
k}} = 2 \, r_k$ {\it per two-mode}. This value is large, but not
maximal. Indeed, when $c_k = 0$ one would have found the maximal
value of the entropy. It is twice the above value, i.e. \ba
S_{{\bf k},\, -{\bf k}}^{max}= S_{{\bf k},\, -{\bf k}}^{inc} = 4
\, r_k\, , \ea or $S^{inc}_{\bf k} = 2 \, r_k$ per mode ${\bf k}$.

It is interesting to notice that the entropy associated with the
effective distribution Eq. \Eqref{Peff} of sine functions equals
that of Eq. \Eqref{rhodec}, up to an arbitrary constant which
arises from the usual ambiguity of attributing an entropy to a
classical distribution. (This ambiguity can be lifted when
introducing $\hbar$ to normalize the phase space integral.) Using
this trick, the entropy associated with Eq. \Eqref{Peff} is
$S_{e\!f\!f}= 2 \, r_k$ for each {\it independent} mode (the
entropy is maximal because the state is Gaussian).
 However, 'for each {independent} mode' here means for each two-mode
 since the mode $-{\bf k }$ is no longer independent
 once the cosines have been neglected, see Eq.
 \Eqref{Aeff}.
 From this equality of entropies
 to leading order in $n_k$,
 we conclude that the quantum density
 matrix which corresponds to Eq. \Eqref{Peff}
 can be taken to be that given by Eq. \Eqref{rhodec}.
 A priori one might have thought that many quantum distributions
 can be associated with
 Eq. \Eqref{Peff}.
 This is not the case when imposing that Gaussianity is preserved
 and that the entropies coincide. Indeed,
 in the high occupation number limit, the entropy is an
 extremely sensitive function of the residual coherence.
 As we shall see in
 Appendix D7,
 this drastically restricts the space of quantum density matrices
 in correspondence with Eq. \Eqref{Peff}.

 Having obtained a lower bound for the entropy from
 quantum considerations, we now provide an upper bound for the
 entropy which could have resulted from processes in the
 inflationary phase.

 \subsection{Maximal entropy and backreaction effects}

The bound follows from the fact that increasing the decoherence
implies increasing the power of the decaying mode, see the power
of the cosines in Eqs. \Eqref{ABvariances} and \Eqref{RRR2}. One
thus obtains an upper bound on the decoherence level when
requiring that the power of the decaying mode be smaller than that
of the growing mode at the onset of the
 radiation dominated era. This requirement follows from the
fact that the r.m.s. value of the primordial fluctuations (of the
Bardeen potential) cannot be much higher than the contribution of
the growing mode from in vacuum because otherwise this would
invalidate the whole framework of linear metric perturbations.
Notice that this type of bound also occurs in the context of the
trans-Planckian question \cite{MB,NP,NPC}, or whenever one is
exploring the upper value of the deviations from the standard
predictions of inflation \cite{cmbParenta}.

 Using Eqs. \Eqref{gdmodes}, \Eqref{RRR2} and \Eqref{RRR3}, and
 expanding the sine and cosine as
 $\cos(k\eta_r) = 1 + O(n^{-3/2})$ and
 $\sin(k\eta_r) = O(n^{-1/4})$
 where we used $n_k \propto (k\eta_r)^{-4} \gg 1$, the condition
 on the power of the decaying mode
 stated above reads $\delta_k \leq O({n_k}^{1/2})$, or
 equivalently
 \ba
   \bar n_k
   \leq  n_k^{3/4} \, , \qquad
   S_{{\bf k},\, -{\bf k}}  \leq
    3 r_k + O(1) \, .
 \ea
 If no further decoherence is added in the radiation dominated
era, this is the maximum amount of entropy stored in the
primordial spectrum. Notice that when evaluated at recombination,
the two-point function is still unaffected by this modification of
the coherence because at that time the decaying mode has still
further decreased. Indeed the relative modifications are then of
the order of $n_k^{-1/2} \sim 10^{-50}$.

 From this analysis we (re-)learn that the dynamical processes
 occurring in the inflationary era which are compatible with a
 linearized treatment of perturbations will, in general,
 leave no significant
 imprint on the spectra at recombination.


 \section{Homogeneous Gaussian distributions with high occupation numbers}

 The purpose of this section is to identify the quantum and classical
 statistical properties of  homogeneous Gaussian distributions
 when the two-mode coherence is not perfect.
 In particular, we shall determine the
 two interesting ranges of decoherence level.
 First there is a narrow range wherein modifications
 of the initial density matrix are so small than the distribution keeps
 quantum properties. Second there is a much wider range
 wherein the distribution, on the one hand,
 has lost its quantum properties so as
 to be interpretable in classical terms,
 but, on the other hand,
 has preserved the temporal coherence of modes.
 We shall only present the results.
 The interested reader will find the technical details
 in Appendix D.

 The most general statistically homogeneous and isotropic Gaussian
 distribution is characterized by its second moments given in Eqs.
 \Eqref{momenta}.
 They are characterized by three real quantities.
 However when focussing on the coherence of the distribution,
 only one parameter
 matters, namely that governing the norm of
 the coherence term.
 In this paper, we shall not discuss the possible modifications
 of the spectrum which are due to a change of the phase of $c_k$.
 Firstly because this change induces no entropy change.
 Secondly because these modifications have been discussed in the context
  of the trans-Planckian question \cite{NP,NPC,MB}. In that case
  indeed  the
  phase shifts result from a change of the Bogoliubov coefficients
  which preserve the purity of the state. Therefore those changes
  are {orthogonal} to the ones considered
  in the present paper.

  A convenient parametrization of the coherence level is provided by
 $\delta_k$ which is defined by
 \ba \label{paramd}
   \abs{c_k}^2 = (n_k + 1)(n_k  - \delta_k) \, .
 \ea
 $\delta_k$ ranges between $0$ and $n_k$.
 They are three characteristic values:
 $\delta_k =0$ obviously corresponds to the pure squeezed state: the
 original {\it in}-vacuum,
 $\delta_k = 1$ corresponds to the critical value above which
 density matrices loose their quantum properties,
 see Appendices D3-6,
 and $\delta_k = n_k$
 corresponds to the thermal case with maximal entropy.



 \subsection{Quantum and classical properties}

 The analysis of Appendix D
 reveals that the statistical properties
 of the distribution
 fall in two categories,
 according to whether they are
 {\it robust} or {\it fragile} when the distribution is slightly perturbed.
 By definition the ``classical'' properties of the distribution, i.e.
 the physical properties observable in the classical regime,
 are the robust ones.

 \subsubsection{Classical properties}

 By robust, we designate the statistical
 properties of
 the original distribution $\hat \rho_{in}$ which
 are preserved in a wide class of partially decohered distributions.
 More precisely, to leading order in $n_k$,
 these properties are unaffected by a
 modification of the coherence term $\Delta |c_k|/|c_k| \ll 1$.
 Thus one has:
 \ba \label{robust}
   {\rm  classical} \qquad \Longleftrightarrow \qquad
   {\rm robust} \qquad \Longleftrightarrow \qquad {\rm unaffected} \,  \,
   {\rm when } \, \, \, \delta_k \ll n_k
   \, .
 \ea
Since the phase of $c_k$ is unchanged, $\delta_k \ll n_k$
guarantees that the power spectrum and the temporal
 coherence of the modes are robust properties of the state.
 As we shall see below, (together with the Gaussianity
which is here exactly preserved) they are the only ones.

 For high occupation numbers, the condition  $\delta_k \ll n_k$
 is a very mild constraint. Indeed, robust
 properties will be lost only when
 the state is close to a thermal state (and thus completely
 incoherent). When starting from a squeezed state, this regime
 can only be reached dynamically in the presence of strong non-linearities.
 Given the amplitudes of primordial fluctuations $\simeq 10^{-5}$,
 this possibility seems excluded in inflation.

 It is interesting to notice that the condition $\delta_k \ll n_k$
 can be interpreted in two complementary ways.
 First, with the mean values. In this case, $\delta_k \ll n_k$
 gives $(\abs{c_k} - n_k) /n_k \ll 1$. This
  coincides with Eq. \Eqref{neqqq}
 which is the
 condition for the two-point function to factorize.
 Alternately, one can consider the distribution itself (see Appendix D.2).
 One gets
 \ba \label{Qtexte}
   \bra{v,\,  {\bf k}}\bra{w,\, - {\bf k}} \hat \rho
   \ket{w,\,  -{\bf k}} \ket{v,\,  {\bf k}}
   \propto \exp\left[ - \frac{\abs{v}^2}{n_k+1} -
   \frac{\abs{w - \bar w}^2}{1+ \delta_k}  \right]
   \, ,
 \ea
 where $\bar w = c_k\, v^*/(n+1)$ designates
 the most probable value of $w$ given $v$.
 It generalizes Eq. \Eqref{meanw}.
 When $\delta_k \ll n_k$
 the spread of $w$ around $\bar w $ is negligible
 with respect to the spread in $v$, the power $(=n_k+1)$.
 Thus the condition
 $\delta_k \ll n_k$
 means that the modes ${\bf k}$ and
 $-{\bf k}$ are still tightly correlated in phase and amplitude,
 as they were in the original distribution, see discussion
 bellow  Eq. \Eqref{vwdistrib}.
 These distributions thus are (to leading order in $n$)
 statistical mixtures of  two-mode coherent states
 $\ket{v} \ket{\bar w}$ as in Eq. \Eqref{rhodec2}.

 \subsubsection{Quantum properties}

 As discussed in Appendix D, the ``quantum features'' of the distribution are
 the following:
 \begin{itemize}
 \item
 The correlations between macroscopically separated
 semi-classical states, see
  discussion after Eq. \Eqref{vwrhov'w'} and Appendix D.3,

 \item
  the existence of a sub-fluctuant
  mode in phase-space, see
 Appendix D.4,

 \item the non-$P$-representability
 of the state, see Appendix D.5,

 \item
 the non-separability of the state
 and the violation of Bell inequalities, see Appendix D.6, and

 \item
 an entropy smaller than that of the minimal scheme, see Eq. \Eqref{Smin}
 and Appendix D.7.
 \end{itemize}

The important fact is that all these properties are erased when
$\delta_k \geq 1$, {\it independently} of $n_k$ when $n_k \gg 1$.
One can thus say
 \ba
  {\rm quantum}  \qquad \Longleftrightarrow \qquad {\rm fragile }
   \qquad \Longleftrightarrow \qquad
   {\rm erased \,\,\, iff}
   \quad \delta_k \geq 1 \, .
 \ea

 Let us emphasize the important lessons.
 The first is that these quantum properties are
 still present when $\delta_k < 1$. Therefore $\delta_k = 1$ is indeed
 the critical value
 from which quantum properties are lost.

 Second,
 the density matrix which corresponds to $\delta_k = 1$
 is that given in Eq. \Eqref{rhodec2}. From Eq. \Eqref{lim},
 one clearly sees that it is the least decohered separable
 distribution.

 Third,
 for $n_k \gg 1$, the quantum properties are very
 sensitive to a slight perturbation of the system since
 $\delta_k = O(1)$
 corresponds to
 a relative change of order $1/n_k$
 of the power spectrum  Eq. \Eqref{powern}.
 Therefore,
 there is a wide range
 of $\delta_k$ (from $1$ to $n_k^{\gamma}$ with $\gamma < 1$) where the robust
 properties are still unaffected (to leading order in $n_k$)
 while the quantum properties are completely erased.

 Fourth, for $\delta_k \leq 1$, the product
 of the residual squeezed spread
 times the residual width of the correlations between
 semi-classical states in the
 orthogonal direction
 stays approximately constant whereas it increases linearly with $\delta_k$
 for $\delta_k \gg 1$, see Eq. \Eqref{interprod} and
 Fig $3$, see also \cite{Morikawa}. This means that
 these two spreads are equivalent expressions of
 the residual quantum properties of the state.

 Fifth, the entropy (which is a measure of the degree of purity of the state)
 is also very sensitive
 to the level of coherence, see Fig. $5$. Therefore
 it can be used
 to parameterize the level of coherence
  and to
  order different distributions, see Appendix D7.

 \subsection{Discussion}

 First, the above results show that
 the following statements are
 {\it equivalent} in the high occupation number limit,
 for decoherence levels obeying $\delta_k \ll n_k$ :
 \begin{itemize}

 \item
 The distribution of the modes ${\bf k}$ and $-{\bf k}$
 is strongly correlated in phase and amplitude,
 where strong means that, given a value 
 $v$ of the amplitude of the mode ${\bf k}$, the conditional probability
 to find 
 ${-{\bf k}}$ with amplitude $w$ is centered around a value
 $\bar w  (v)$ with a
 spread $=1 + \delta_k \ll n_k$, see Eq. \Eqref{Qtexte}.

  \item
  All realizations of
 the modes contributing to the two-point function
 have the same temporal
 phase. This phase is fixed by $\arg(\bar w)$ of Eq.
 \Eqref{Qtexte} and is not affected by the level of decoherence.
 Indeed the modifications of $\arg(\bar w)$ preserve the
 quantum purity  of the distribution and are orthogonal
 to those considered here.
 The absence of phase shift
 guarantees  that
 the decaying mode stays subdominant for $\delta_k \ll n_k$, in the sense that
 its power
 $(= \delta_k/2+ O(1/n))$ stays $\ll n_k$, see Eq. \Eqref{RRR2}.

 \item
  The two-point function factorizes in a product of two growing modes.

\end{itemize}

 Second, both the power spectrum and the temporal coherence
 are fairly robust, in the sense
 that they are lost only for $\delta_k \simeq n_k$,
 i.e. through a significant regeneration of the decaying mode.
  They are the statistical properties of the state
 which can be interpreted classically.
 The fact that the criterion of robustness singles out
 certain statistical properties
 should be conceived as a {\it definition} of
 the classical (statistical)
 information contained in the quantum density matrix,
 in agreement with Gottfried's analysis \cite{gottfried}.

In the case where $\delta_k < 1$, the distribution cannot be
interpreted in classical terms. Nevertheless, when probing (or
observing) the momenta of this distribution with a relative
uncertainty larger than $1/n_k$, the above robustness prevents the
existence of signals telling us that some quantum properties are
still present. Indeed, quantum properties of Gaussian
distributions can {\it only} be put into evidence
in two cases. Either one must be able
to measure the momenta of distribution $\Eqref{momenta}$
with a precision higher than $1/n_k$. This amounts to be able
to measure the amplitude of the sub-fluctuant mode
and hence that of the decaying mode, see Eq. \Eqref{RRR2}.
Or one must have a set of
non-quadratic operators. A first exemple of such operator
giving rise to violations of Bell's inequalities
can be found in \cite{Wodkievicz}, see also
Appendix D6; a second is discussed in \cite{CP4}.
The consistency of these two alternatives
is understood when noticing that the measures involved
in the second case amount to distinguish configurations
with a precision higher than $1/n_k$.


These considerations imply that the non-linear processes giving
rise to structure formation will also be insensitive to the
existence of a residual degree of quantum coherence. (In this we
disagree with the claim \cite{HuCalzetta} according to which only
the decohered part of the power spectrum will participate to
structure formation).
The question of the residual quantum coherence (i.e. the value of
$\delta_k$ if $\delta_k < 1$) seems therefore of purely academic
interest.




 \section{Conclusion}

 In this paper we
 analyzed  partially decohered distributions
 susceptible to describe the primordial fluctuations
 in inflationary scenarios.

 First we have
 identified the parameter governing the level of decoherence
 and computed the associated entropy, the residual value of the
 squeezing, and the residual coherence encoded in
 off-diagonal matrix elements.

 Second, we showed
 that the
 correlations of modes in the
 initial distribution and its decohered versions,
 which is at the origin of the temporal coherence of the
 mode at horizon re-entry,
 is properly taken into account by considering superpositions
 of 'two-mode coherent states'.

 Third, we discussed how, in the large occupation number limit, the
 fragility of quantum properties of the distribution to
 small modifications
 lead to a critical decoherence scheme above which all quantum
 properties are erased. The associated entropy is very large
 (and equal to $1/2$ the thermal entropy) even though
 the relative change of the
 power spectrum is only $1/n_k$.
 For $\delta_k > 1$, the remaining statistical properties can
 be interpreted classically.

 The open question concerns the calculation
 of the residual degree of coherence (the value of $\delta_k$).
 This is a difficult  question which
 calls for further work.
 During the inflationnary period,
 the dynamics is indeed not trivial because of the conjunction of
 mode amplification and non-linear effects.


\vskip .5 truecm

 {\bf Acknowledgments:}
 We are grateful to Serge Massar for having
 explained to us the meaning of decoherence by drawing Fig. 1 many
 years ago in Brussels. We are also grateful to Dani Arteaga,
 Claus Kiefer, Jihad Mourad and Alexei Starobinsky for useful remarks.


\begin{appendix}

\section{Coherent states}

This appendix aims to present the properties which we shall use in
the body of the manuscript. For more details, we refer to
\cite{Glauber63a,Glauber63b,Zhang90}.

Coherent states (of a real oscillator) can be defined as
eigenstates of the annihilation operator:
 \ba \label{DefCoh1}
   \hat a \ket{v} = v \ket{v} \, ,
 \ea
where $v$ is a complex number. In Fock basis it is written as
 \ba \label{DefCoh2}
   \ket{v} = e^{-\frac{\vert v \vert^2}{2}}  \sum_{n=0}^{\infty}
   \frac{v^n}{\sqrt{n!}} \ket{n} \, ,
 \ea
where the exponential prefactor guarantees that the state is normalized
to unity $\langle v \vert v \rangle=1$.
They are also obtained by the action of a displacement operator on the vacuum :
 \ba \label{displacement}
   \ket{v} = \hat D(v) \ket{0} = e^{v^* \hat a - v \hat a^{\dagger}}
   \ket{0} \, .
 \ea

The first interesting property of coherent states is that they correspond
to states with a well defined complex amplitude $v$.
Indeed, by definition \Eqref{DefCoh1},
the expectation values of the annihilation and creation operators are
 \ba
   \bra{v} \hat a \ket{v} = v \, , \quad
   \bra{v} \hat a^{\dagger} \ket{v} = v^{*}
   \, .
 \ea
Thus the mean occupation number is
 \ba \label{occupnumber}
   \bra{v} \hat a^{\dagger} \hat a \ket{v} = \abs{v}^2 \, .
 \ea
It is to be also stressed that the variances vanish:
 \ba \label{nullvariance}
   \Delta \hat a^2 = \bra{v} \hat a^2 \ket{v} - \bra{v} \hat a \ket{v}^2 = 0 \, ,
   \quad \Delta \hat a^{\dagger \, 2} = \bra{v} \hat a^{\dagger \, 2} \ket{v} -
   \bra{v} \hat a^{\dagger} \ket{v}^2 =  0 \, .
 \ea

 From these properties one sees that the expectation values of the
 position and momentum operators (in the Heisenberg picture)
 \ba
   \hat q(t)=\sqrt{\frac{\hbar}{2\omega}}\left(\hat a e^{-i \omega t} +
        \hat a^{\dagger} e^{i \omega t}\right) \, , \qquad
   \hat p(t)=-i\sqrt\frac{\hbar\omega}{2}(\hat a e^{-i \omega t} -
                \hat a^{\dagger}e^{i \omega t})
   \, , \nonumber
 \ea
 are
 \ba \label{cltrajincohstate}
   \bar q(t) &=& \bra{v} \hat q(t) \ket{v} =
   \sqrt{\frac{\hbar}{2\omega}}( v e^{-i \omega t} + v^* e^{i \omega t})
            = \sqrt{\frac{2\hbar}{\omega}} \abs{v} \cos(\omega t - \phi_v)
   \, , \nonumber \\
   \bar p(t) &=& \bra{v} \hat p(t) \ket{v} =
   -i\sqrt{\frac{\hbar\omega}{2}} ( v e^{-i \omega t} - v^* e^{i \omega t})
            = -\sqrt{2\hbar\omega} \abs{v} \sin(\omega t - \phi_v)
        = \partial_t \bar q(t) \, .
 \ea
We have used the polar decomposition $v=\abs{v} e^{i \phi_v}$.
These expectation values have a well defined amplitude and phase
and follow a classical trajectory of the oscillator. This is due
to the ``stability'' of coherent states under the evolution of the
free Hamiltonian $2 H_0= p^2 + \omega^2 q^2$: if the state is
$\ket{v}$ at time $t_0$, one immediately gets from \Eqref{DefCoh2}
that at a later time $t$, the state is given by $\ket{v(t)} =
\ket{ve^{-i\omega (t-t_0)}}$. Notice that the variances of the
position and the momentum are
 \ba
   \Delta \hat q^2 = \frac{\hbar}{2\omega} \, , \qquad
   \Delta \hat p^2  = \frac{\hbar\omega}{2} \, .
 \ea
 They minimize the Heisenberg uncertainty relations and are
time-independent. Hence, in the phase space $(q,p)$, a coherent
state can be considered as a unit quantum cell $2\pi \hbar$ in
physical units (see also \Eqref{measure} for the measure of
integration over phase space) centered on the classical position
and momentum of the harmonic oscillator $(\bar q(t),\bar p(t))$.
In the large occupation number limit $\abs{v}\gg 1$, coherent
states can therefore be interpreted as classical states since
$\Delta \hat q / \sqrt{\bar q^2+ \bar p^2/\omega^2} = \Delta \hat
p / \sqrt{\omega^2\bar q^2+ \bar p^2} = 1/2\abs{v}$. This is a
special application of the fact that coherent states can in
general be used to define the classical limit of a quantum theory,
see \cite{Zhang90} and references therein.

One advantage of coherent  states \cite{Glauber63a} is that the
calculations of Green functions resembles closely to those of the
corresponding classical theory (i.e. treating the fields not as
operators but as c-numbers) provided either one uses normal
ordering, or one considers only the dominant contribution when
$\abs{v}\gg 1$. We compute the Wightman function in the coherent
state $\ket{v}$
 \ba
   \widetilde G_v(t,t') &=& \bra{v} \hat q(t) \hat q(t') \ket{v} \nonumber \\
           &=& \langle :\hat q(t)\hat q(t') : \rangle_v +
       \frac{\hbar}{2\omega} e^{i\omega(t-t')} \, ,
 \ea
where we have isolated the contribution of the vacuum.
The normal ordered correlator is order $\abs{v}^2$:
 \ba \label{cohgreen}
   \langle :\hat q(t)\hat q(t') : \rangle_v &=&
   \frac{\hbar}{\omega} \,
   Re\left[\langle \hat a^2 \rangle_v e^{-i\omega(t+t')} +
    \langle  \hat a^{\dagger} \hat a \rangle_v e^{i\omega(t-t')} \right]
    \nonumber  \\
   &=& \frac{2\hbar}{\omega} \vert v \vert^2 \cos(\omega t - \phi_v)
     \cos(\omega t' -\phi_v)   = \bar q(t) \,  \bar q(t') \, .
 \ea
We see that the perfect coherence of the state, namely
$\abs{\langle \hat a  \hat a  \rangle_v }
= \langle \hat a^\dagger  \hat a  \rangle_v  $
is necessary to
combine the contributions of the diagonal and the interfering term so as to
bring the time-dependent classical position $ \bar q(t)$ in
Eq. \Eqref{cohgreen}.

The wave-function of a coherent state in the coordinate
representation is given by
 \ba \label{CohWaveFunct}
    \psi_{v}(q) = \left(\frac{\omega}{\pi\hbar}\right)^{1/4}
    \exp\left(-\frac{\omega}{2\hbar}(q-\bar q)^2 -i \frac{\bar p q}{\hbar} \right)
    \, ,
 \ea
 where $v=(\omega \bar q + i \bar p)/\sqrt{2\omega\hbar}$. This
follows from the definition $\bra{q} \hat a \ket{v} = v \langle q
\vert v \rangle$. From this equation one notes that two coherent
states are not orthogonal. The overlap between two coherent states
is
 \ba \label{overlap}
   \langle v \vert w \rangle = \exp\left(v^{*} w - \inv{2}\abs{v}^2 -
   \inv{2}\abs{w}^2 \right)  \, .
 \ea
 Nevertheless they form an (over)complete basis of the Hilbert
space in that the identity operator in the coherent state
representation $\{ \ket{v} \}$ reads
 \ba \label{coh1}
   {\bf 1} = \int\!\!\frac{d^2v}{\pi} \, \ket{v} \bra{v} \, .
 \ea
The measure is
 \ba \label{measure}
   \frac{d^2v}{\pi} = \frac{d({\mathrm Re} v) d({\mathrm Im} v)}{\pi} =
   \frac{d\bar q d\bar p}{2\pi \hbar} \, .
 \ea
The representation of identity can be established by calculating
the matrix elements of both sides of
the equality in the coordinate representation $\{ \ket{q} \}$,
with the help of \Eqref{CohWaveFunct}.


\section{Application of Zurek \& al. results to the cosmological case}

In this appendix we show that Eq. \Eqref{rhodec} is 
the minimally decohered distribution in the sense of \cite{Zurek93}.
To this end, we decompose the complex mode $\hat \phi_{\bf k}$
into two real oscillators $\hat \phi_{1}$ and $\hat \phi_{2}$
given by its real and imaginary parts. Since the two-mode
Hamiltonian is quadratic and Hermitian, it splits into the sum of two identical
one-mode oscillator Hamiltonians for $\hat \phi_{1}$ and $\hat
\phi_{2}$ separately. Notice that the annihilation operators of
these two real oscillators,
\ba
\hat a_1 = \left(\hat a_{\bf k} +
\hat a_{-\bf k}\right)/\sqrt{2} , \, \quad  \hat a_2 = -i\, \left(\hat
a_{\bf k} - \hat a_{-\bf k}\right)/\sqrt{2}, \,
\label{artop}
\ea
 mix ${\bf k}$ and
$-{\bf k}$ annihilation operators. Hence they can easily enforce
correlations between these two modes. However, for our decoherence
procedure to be valid, as noticed in \cite{KiefPolStar98a},
it is important that the interactions
 do not mix $\hat \phi_{1}$ and $\hat \phi_{2}$,
 so that they
 do not break the homogeneity.

A two-mode squeezed state $\kett{0 in, {\bf k}}$ can always be
written as the tensorial product of the two one-mode squeezed
states \cite{schum}. In our case, the one-mode squeezed states are
those of the oscillators $1$ and $2$ because the Hamiltonian
separates. Thus we have
 \ba \label{factor12}
   \kett{0 in, {\bf k}} = \ket{0 in, 1} \otimes \ket{0 in, 2} \, .
 \ea
 The one-mode squeezed states are governed by the same parameter $z/2$ :
 \ba
   \ket{0 in, 1} = \frac{1}{\sqrt{\abs{\alpha}}} \sum\limits_{n=0}^{\infty}
   \left( \frac{z}{2} \right)^n \frac{\sqrt{2n !}}{n!} \ket{2 n,\, 1} \, .
 \ea
 The same expression holds for the ket $\ket{0 in, 2}$.

 The overlap of this one-mode squeezed state with a one-mode coherent state is
 \ba
   \langle v ,\, 1 \vert 0 in, 1  \rangle =
   \frac{1}{\sqrt{\abs{\alpha}}} \exp\left(- \abs{v_1}^2/2+ zv_1^{*\, 2}/2 \right) \, .
 \ea

According to \cite{Zurek93}, when taking small interactions into
account, the density matrix of a one-mode squeezed state will
preferably decohere into the mixture
 \ba
   \hat \rho_{red,\, 1} =  \int\!\!\frac{d^2 v_1}{\pi} \,
   {\cal P}_1(v_1)  \,
   \ket{v_1} \bra{v_1} \, .
 \ea
 where the statistical weight is given by the probability to find a coherent
 state starting from the initial state: 
 \ba \label{proba1}
    {\cal P}_1(v_1) = \abs{\langle v ,\, 1 \vert 0 in, 1  \rangle }^2 &=&
    \frac{1}{\abs{\alpha}} \exp\left(- \abs{v_1}^2 + Re(zv_1^{*\, 2}) \right)
   \, .
 \ea
 It should be noticed that this choice of the probability is not unique.
 The procedure adopted here
 consists in taking the so-called $Q$-representation of the original density
 matrix
 (see Appendix D.1 for its definition)
 and to treat it as a $P$-representation to define the
  the decohered matrix density. A slightly more coherent
  distribution is given at the end
  of Appendix D5. Since the ambiguity results from
  the fact that coherent states are
   overcomplete,
no differences show up to leading order in $n_k$.
   Hence, the various distributions belong to the same class:
   they have the same entropy, see Appendix D7.


To complete the proof we need to show that product of two one-mode
coherent states $1$ and $2$ is also the product of a coherent
state for the ${\bf k}$ and $-{\bf k}$ modes. This is the case:
 \ba
   \ket{v_1}\otimes\ket{v_2} &=&
   \hat D_1(v_1) \hat D_2(v_2) \ket{0 in, 1} \otimes \ket{0 in, 2}
   \, , \nonumber \\
   &=& \hat D_{\bf k}(v) \hat D_{-{\bf k}}(w)
   \ket{0 in, {\bf k}} \otimes \ket{0 in, -{\bf k}}
   \, , \nonumber \\
   &=&  \ket{v, {\bf k}} \otimes \ket{w, -{\bf k}} \, ,
 \ea
 where the displacement operator $\hat D$ is defined at
 Eq. \Eqref{displacement}, and
 where the amplitudes are related by
 \ba
   v=\frac{v_1+iv_2}{\sqrt{2}} \, , \qquad w=\frac{v_1-iv_2}{\sqrt{2}} \, .
 \ea
Finally, the product of the probabilities \Eqref{proba1} gives the probability
Eq. \Eqref{vwdistrib}. Performing the change of variables of integration
from $(v_1,\, v_2)$ to $(v,\, w)$ completes the proof.




\section{The covariance matrix}



 Let us consider a given pair of modes $({\bf k}, \, - {\bf k})$
 in a Gaussian state \Eqref{momenta}. We prove the
 relations (\ref{map}-\ref{tilde n}). This is best done with the
 covariance matrix of the state. We also define the sub- and
 super-fluctuant modes.

 \subsection{Definition of the covariance matrix}

To conform ourselves to the usual definition of the
covariance matrix \cite{Simon00},
 we introduce canonically conjugated and
 {\it hermitian} 'position' and 'momentum'
 variables for each mode, i.e.
 $\hat a_{{\bf k}} = (\hat 
   q_{{\bf k}} +
 i \hat p_{{\bf k}})/\sqrt{2 }$,
 and similarly
 $\hat a_{-{\bf k}} = ( \hat q_{-{\bf k}} +
 i\hat p_{-{\bf k}})/\sqrt{2 }$.
 All the 
 information of the Gaussian state
 is encoded in the covariance matrix.
 In terms of the 
 $4$ vector
 \ba
   {\bf \zeta} =
   \left(
     \begin{array}{c}
     \hat q_{{\bf k}} \\
     \hat p_{{\bf k}} \\
     \hat q_{-{\bf k}} \\
     \hat p_{-{\bf k}}
     \end{array}
   \right) \, ,
 \label{zeta}
 \ea
 the covariance matrix has
 the matrix elements
 $C_{ij}= 
 {\rm Tr}\left(
 \hat \rho \left\{ \, \hat {\bf \zeta}_i , \,
 \hat {\bf \zeta}_j^{T} \right\} \right)$,
 where $\left\{ \, , \, \right\}$ is the
 Weyl ordered product, i.e.
 the anticommutator divided by 2.
 For the states with variances \Eqref{momenta}, one has
 \ba \label{covmatrix}
   C =
   \left(
     \begin{array}{c c c c}
      n + \frac{1}{2} & 0 &  c_r & c_i  \\
      0 & n + \frac{1}{2} & c_i & -  c_r  \\
      c_r & c_i & n + \frac{1}{2} & 0 \\
      c_i & - c_r  & 0 & n + \frac{1}{2}
     \end{array}
   \right) \, ,
 \ea
 where $c_r = Re(c) = - \vert c \vert \cos 2\theta$,
 and $c_i = Im(c) = - \vert c \vert \sin 2\theta$,
 see Eq. \Eqref{epsilontheta}.
 This covariance matrix is not bloc diagonal
 since the modes ${\bf k}$ and $- {\bf k}$ are correlated.
 As we shall see in Appendix D6, this will lead to
 the notion of non-separability.

 \subsection{The sub- and super-fluctuant modes}

 It is appropriate to return to complex modes to express the
 eigenmodes and the eigenvalues of the covariance matrix \Eqref{covmatrix}.
 They
  are given by the rotated  modes:
 \ba \label{rotatedvar}
   \Phi_{\bf k} = \frac{1}{\sqrt{2 }} (\hat a_{\bf k}e^{-i\theta} +
   \hat a_{-\bf k}^{\dagger}e^{+i\theta}) \, , \qquad
   \Pi_{\bf k} = -i \frac{1}{\sqrt{2}} (\hat a_{\bf k}e^{-i\theta} -
   \hat a_{-\bf k}^{\dagger}e^{+i\theta})
   \, .
 \ea
 Their expectation values are 
 \ba \label{eigenvalues}
  \langle \{ \Phi_{\bf k} , \, \Phi^\dagger_{\bf k} \} \rangle =
   \left( n+\frac{1}{2} \right) + \abs{c} \, , \quad
   \langle \{ \Pi_{\bf k} , \, \Pi^\dagger_{\bf k} \} \rangle =
   \left( n+\frac{1}{2} \right) - \abs{c} \, ,  \quad
   \langle \{ \Phi_{\bf k} , \, \Pi^\dagger_{\bf k} \} \rangle = 0
  \, . \nonumber \\
 \ea
The first two r.h.s. expressions 
give the eigenvalues of \Eqref{covmatrix}.
They are twice degenerate because of homogeneity.
In the case of pair creation from the vacuum,
$n=\sh^2(r)$ and $\abs{c} = \ch r \, \sh r$,
they reduce respectively to $e^{2r}/2$ and $e^{-2r}/2$.
 They are therefore
 called the super-fluctuant and sub-fluctuant modes respectively.
 It is worth noticing
 that their product gives $1/4$ as in the vacuum. This is
 the expression of the purity of the state. 

 \subsection{Proof of Eq. \Eqref{map}}

An efficient way to prove
Eq. \Eqref{map} and
Eq. \Eqref{tilde n} consists in using
the 
the real and the imaginary parts of $\phi_{\bf k}$,
see Eq. \Eqref{artop}.
 The passage from $({\bf k},\, -{\bf k}) \mapsto (1,\, 2)$
 is performed by making
 a {\it global} rotation $R_{g   }$ of angle $\pi/4$.
 By global we mean
 a $4 \times 4$ transformation  mixing
 ${\bf k}$ and $-{\bf k}$ sectors. Explicitly, one has
 \ba \label{U}
   \left(
     \begin{array}{c}
    \sqrt{k} \, \hat \phi_{1}  \\
    \hat \pi_{1}/\sqrt{k} \\
    \hat \pi_{2}/\sqrt{k} \\
    - \sqrt{k} \,\hat \phi_{2}
     \end{array}
   \right) =
   \frac{1}{\sqrt{2}}
   \left(
     \begin{array}{c c c c}
      1 & 0 & 1 & 0 \\
      0 & 1 & 0 & 1 \\
      -1 & 0 & 1 & 0 \\
      0 & -1 & 0 & 1
     \end{array}
   \right)
   \left(
     \begin{array}{c}
     \hat q_{{\bf k}} \\
     \hat p_{{\bf k}} \\
     \hat q_{-{\bf k}} \\
     \hat p_{-{\bf k}}
     \end{array}
   \right) \, .
 \ea
 The covariance matrix becomes bloc diagonal:
 \ba \label{covmatrixfictious}
   C \mapsto C_{(1,2)} = R_{g} C R_{g}^T =
   \left(
     \begin{array}{c c c c}
      n_{k} + \frac{1}{2} + c_r & c_i & 0 & 0 \\
      c_i & n_{k} + \frac{1}{2} - c_r & 0 & 0 \\
      0 & 0 & n_{k} + \frac{1}{2} - c_r & c_i \\
      0 & 0 & c_i & n_{k} + \frac{1}{2} + c_r
     \end{array}
   \right) \, .
 \ea
 The state is a tensor product
 $\hat \rho = \hat \rho_1 \otimes \hat \rho_2$
 and the  matrices $\hat \rho_1$
 and $\hat \rho_2$ coincide. These properties
 follow from the homogeneity of the state.


 The transformation Eq. \Eqref{map} amounts to bring $C$ under the
 form
 \ba
   C = M T  M^T \, , \qquad
   T =  \left( \bar n + \frac{1}{2}\right) {\bf 1} \, ,
 \ea
 where the matrix $T$ is the covariance matrix of
 the product of the two thermal
 density matrices $\hat \rho_{th,\, 1} \otimes \hat \rho_{th,\, 2}$
 in Eq. \Eqref{map}. The matrix $M$ is the product of three 
 special matrices 
 \ba
   M = S_{loc}(\kappa) \, R_{loc}(\varphi) \, 
   R_{g   } \, .
 \ea
 First, the global rotation $R_{g   }$ of \Eqref{U}.
 Second, the 
 product of
 $2 \times 2 $, i.e. {\it local},
 rotations, $R_{loc}(\varphi) = R(\varphi) \otimes  R(\varphi)$,  where
 \ba \label{SR}
    R(\varphi) &=&
    \left(
     \begin{array}{c c}
     \cos \varphi & \sin \varphi \\
     -\sin \varphi & \cos \varphi
     \end{array}
    \right)  \, , \qquad
    \varphi = \frac{1}{2} \arg(c) = \frac{\pi}{2} + \theta
    \, ,
 \ea
 brings $C_{(1,\,2)}$ into a diagonal form:
 ${\rm diag}(\lambda_+,\, \lambda_-,\, \lambda_-,\, \lambda_+)$
 with $\lambda_\pm = n+1/2 \pm \abs{c}$,
 see \Eqref{eigenvalues}.
 Third two local squeezing $S(\kappa) \otimes S(-\kappa)$
 \ba \label{S}
    S(\kappa) =
    \left(
     \begin{array}{c c}
     e^{\kappa/2} & 0 \\
     0 &  e^{-\kappa/2}
    \end{array}
    \right)  \, , \qquad
    \th \kappa = - \abs{c}/(n + 1/2)
    \, ,
 \ea
 rescale the eigenvalues to a common value $\bar n$.
 The latter is fixed by the conservation of the determinant:
 \ba
   \left(\bar n + \frac{1}{2}\right)^4 =
   {\rm det} \, C = \left( (n + \frac{1}{2})^2 - \abs{c}^2 \right)^2
   \, .
 \ea


\section{General description of homogeneous Gaussian states}

 We review the statistical properties of two-mode (or
 bi-party) Gaussian states. They have received much attention
 because of their importance in the contexts of Quantum Optics and
 Quantum Information, see for instance \cite{Wodkiewiczlecture} for
 a comprehensive review.

 Since in cosmology one deals with statistically homogeneous
 and isotropic distributions,
 one needs to consider only a
 sub-class of Gaussian states. These
 symmetries considerably simplify
 the discussion since the
 density matrices are characterized by
 only three real parameters
 given in Eqs. \Eqref{momenta}.
 It should be noticed that the squeezed
 states \Eqref{inoutvac}
 (including the case of zero squeezing, i.e. the vacuum)
 are the only pure states. They are characterized by only two
 independent parameters.
 This is because
 squeezed vacuum states minimize
 the Heisenberg uncertainty relations, see Eqs. \Eqref{positivity}
 and \Eqref{Heis}.
 The thermal state is
 the most decohered distribution and corresponds to $c=0$.
 As in the body of the text,
 we shall parameterize the level of decoherence by $\delta$
 defined by
 \ba \label{deltafine}
   \abs{c}^2 = (n+1)(n - \delta) \, ,
 \ea
and leave the phase of $c$ unchanged.
 Nevertheless,
 for completeness, we consider in Appendix D.8 the effects of a
 smearing of the {phase} of $c$.


 \subsection{The $Q$-representation of a density matrix}

 To proceed, it is convenient to work with a representation of the
 density matrix $\hat \rho$ in phase-space. We choose the so-called
 $Q$-representation $Q_{\rho}$.
 We recall that it is well defined
 for every hermitian, positive definite operator (density matrix).
 For a two-mode state, $Q_{\rho}$ is the expectation value of
 $\hat \rho$ in a pair of coherent states
 \ba \label{defQrepr}
   Q_{\rho}(v,w) = \bra{v, \, {\bf k}}\bra{w, \, -{\bf k}} \hat \rho_{\bf k,\, -{\bf k}}
   \ket{w, \, -{\bf k}}\ket{v, \, {\bf k}} \, .
 \ea
 It is a function over phase space with the convention
 $v=(k \phi_{\bf k} + i \pi_{\bf k})/ \sqrt{2k}$, and
 $w=(k \phi_{-{\bf k}} + i \pi_{-{\bf k}})/ \sqrt{2k}$, see
 Appendix A.
 It is remarkable that although \Eqref{defQrepr} is an expectation value, the
 knowledge of
 the function $Q_{\rho}(v,w)$ is completely equivalent to that of
 $\hat \rho$. In particular the information about the off-diagonal matrix
 elements $\bra{v}\bra{w} \hat \rho \ket{w'}\ket{v'},\, v\neq v',\,
 w \neq w'$ is encoded in \Eqref{defQrepr} \cite{Glauber69,Nussenzveig}.
 This is due to the
 redundant (overcomplete) character of the basis of coherent
 states
 \footnote{Let us work with one-mode states for simplicity.
 Consider the expansion in the Fock representation of
 a (bounded) operator $\hat O = \sum\limits_{n,m=0}^{\infty}
   \ket{n} O_{nm} \bra{m}$. Using
 \Eqref{DefCoh2}, one sees that the function
 \ba
   {\cal O}(v^*, v') =
   \exp{\left[\frac{1}{2}\left(\abs{v}^2 + \abs{v'}^2 \right)
   \right]} \bra{v} \hat O \ket{v'} =
   \sum\limits_{n,m=0}^{\infty}
   O_{nm} \frac{(v^*)^n (v')^m}{\sqrt{n!m!}}
   \, , \nonumber
 \ea
 is an entire analytic function of both variables $(v^*, v')$.
 A theorem for functions of several complex variables states that if
 ${\cal O}(v^*, v')$ vanishes on the line $v' = v^*$, then it
 vanishes identically \cite{Nussenzveig}.
 Consider then two density matrices
 $\hat \rho_1$ and $\hat \rho_2$ with identical diagonal
 matrix elements in the basis of coherent states. Setting,
 $\hat O = \hat \rho_1 - \hat \rho_2$, and using the previous
 theorem, one concludes that $\hat \rho_1 = \hat \rho_2$.
 }, see Appendix A.

 In addition, for Gaussian states, the knowledge of the expectation values
 \Eqref{momenta} is all one needs to write the $Q$-representation
 of the density matrix. The general method of characteristic
 functions is given in every textbook of
 Quantum Optics, see for instance
 \cite{Glauber69,Nussenzveig,Wodkiewiczlecture}.
 The case of homogeneous distributions can be worked out by hand,
 and one finds
 \ba \label{Qrepr}
   Q_{\rho}(v,w) = \frac{1}{\Delta} \exp\left[-\frac{1}{\Delta}
   \left( (n+1)(\abs{v}^2 + \abs{w}^2 ) - 2 Re(c^*vw) \right) \right]
   \, ,
 \ea
 where
 \ba
   \Delta = (n+1)^2 - \abs{c}^2 = (n+1)(1 + \delta )\, .
 \ea
 One verifies that \Eqref{Qrepr} gives back  \Eqref{vwdistrib}
 when considering two-mode squeezed vacuum states, i.e. $\Delta = n+1$.

 A necessary and sufficient condition for \Eqref{Qrepr}
 to be the expectation value of a
 density matrix (a positive hermitian operator) is
 \ba \label{positivity}
   n(n+1) - \abs{c}^2 \geq 0 \, .
 \ea
 This is easily seen from the fact that $\hat \rho$ is unitarily
 equivalent to a tensor product of thermal states, see Eq.
 \Eqref{map} and Appendix C. From \Eqref{tilde n},
 the positivity condition $\bar n \geq 0$
 gives \Eqref{positivity}. It should be stressed that \Eqref{positivity}
 {\it is} the Heisenberg's uncertainty relations
 \footnote{We recall that the general form of the uncertainty
 relations is, for all hermitian operators $\hat A$ and $\hat B$,
 $\Delta \hat A^2 \Delta \hat B^2 \geq
 \abs{\langle \left[\hat A , \, \hat B \right]\rangle}^2/4$.
 To get \Eqref{Heis} use
 $\hat A = \hat \phi_{\bf k} \hat \phi_{\bf k}^{\dagger}$ and
 $\hat B = \hat \pi_{\bf k} \hat \pi_{\bf k}^{\dagger}$.
 },
 \ba \label{Heis}
   \langle \hat a_{{\bf k}} \hat a_{{\bf k}}^{\dagger} \rangle
   \langle \hat a_{-{\bf k}}^{\dagger} \hat a_{-{\bf k}} \rangle \geq
   \langle \hat a_{{\bf k}} \hat a_{-{\bf k}} \rangle
   \langle \hat a_{{\bf k}}^{\dagger} \hat a_{-{\bf k}}^{\dagger} \rangle
   \,  .
 \ea
 Notice that the lower bound is reached for the squeezed states
 \Eqref{inoutvac}.

 Incidentally, the positivity condition \Eqref{positivity}
 furnishes a clear way of distinguishing classical phase-space
 distributions (i.e. Probability Distributions Functions, the states
 being points in phase-phase),
 and phase-space {\it representations} of quantum states. Were
 $Q_{\rho}(v,w)$ be a PDF, the positivity criterion would have
 been $\Delta
 \geq 0$
 which is less restrictive
 than \Eqref{positivity}.

 The off-diagonal matrix elements are \cite{Glauber69,Wodkiewiczlecture}
 \ba \label{offdiag}
   \bra{v}\bra{w} \hat \rho \ket{w'}\ket{v'} =
   &&\frac{1}{\Delta} \exp\left[-\frac{1}{2} \left(\abs{v}^2 + {\abs v'}^{2}
   + \abs{w}^2 + {\abs w'}^{2}\right) \right] \nonumber \\
   &&\times
   \exp\left[\left(1-\frac{n+1}{\Delta}\right)
   \left( v^* v' + w^* w'\right) + \frac{1}{\Delta} \left(
   c^* v'w' + c v^{*} w^{*} \right) \right] \, ,
 \ea
 and the density matrix itself
 is
 \ba \label{rho}
   \hat \rho &=& \frac{1}{\Delta} \exp\left[ \frac{c}{\Delta}
   \hat a_{\bf k}^{\dagger} \hat a_{-{\bf k}}^{\dagger} \right]
   \, : \exp\left[ - \frac{n+1}{\Delta}
   \left( \hat a_{\bf k}^{\dagger} \hat a_{\bf k} +
   \hat a_{-{\bf k}}^{\dagger} \hat a_{-{\bf k}} \right) \right]
   : \,  \exp\left[ \frac{c^*}{\Delta}
   \hat a_{\bf k} \hat a_{-{\bf k}} \right]
   \, , \nonumber
 \ea
 where ``$: \, : $`` stands for normal ordering.
This completes the proof that the knowledge of $c$ and $n$
allows to reconstruct the {\it quantum} distribution.

 \subsection{Factorisability of the two-point function, the
 classical statistical properties of the state, and two-mode
 coherent states}

 The function $Q_{\rho}$ can be re-written as the product
 of a
 marginal probability $Q_v(v) = \int\!\!\frac{d^2 w}{\pi} \, Q(v,w)$ and
 a conditional probability $Q_{w|v}(w)$
 \ba \label{Q}
   Q_{\rho}(v,w) &=& Q_{v}(v) \times Q_{w|v}(w) \, , \nonumber \\
   &=& \frac{1}{n+1} \exp\left[ - \frac{\abs{v}^2}{(n+1)} \right]
   \times   \frac{1}{1+ \delta} \exp\left[ -
   \frac{\abs{w - \bar w(v)}^2}{1+\delta}  \right]
   \, .
 \ea
 Both $Q_v$ and $Q_{w|v}$ are normalized.
 The quantity $\bar w(v) = cv^*/(n+1)$ is the
 conditional mean value of $w$ given $v$.
 One recovers Eq. \Eqref{meanw}
 for a squeezed vacuum state.
 The conditional width around this conditional mean value is
 \footnote{The means and variances can either be read from \Eqref{Q}, or calculated using
 $v$ and $w$ as random variable whose PDF is \Eqref{Q}. One immediately
 gets
 \ba
   &&\av{v} =
   \int\!\!\frac{d^2 v}{\pi}\frac{d^2 w}{\pi} \, v Q_{\rho}(v,w) =
   0 \, , \qquad
   \av{\abs{v}^2} =
   \int\!\!\frac{d^2 v}{\pi}\frac{d^2 w}{\pi} \, \abs{v}^2
   Q_{\rho}(v,w) =  n+1
   \, . \nonumber
 \ea
 Similar equations hold for the averages $\av{w} = 0$ and
 $\av{\abs{w}^2}=n+1$ since \Eqref{Qrepr} is symmetric in $v$ and $w$.
 The {\it conditional} means and variance of $w$ are calculated
 with the conditional PDF,
 \ba
   \langle w \rangle = \int\!\!\frac{d^2 w}{\pi} \, w
   Q_{w|v}(w) = \bar w(v) \, , \qquad
   \Delta_v w^2 =  \langle \abs{w}^2 \rangle - \abs{\bar w}^2 =
   \int\!\!\frac{d^2 w}{\pi} \,
   \left(\abs{w}^2 - \abs{\bar w}^2 \right) Q_{w|v}(w)  = 1+\delta   \, . \nonumber
 \ea
 Notice that this spread is independent of the value of $v$. This
 guarantees the temporal coherence in the mean
 $\Delta w^2 = \av{\abs{w}^2} - \av{\abs{\bar w}^2} = 1 + \delta$,
 see discussion below \Eqref{strongpeak}.
 For subtleties concerning operator ordering  using
 phase-space representations of the density matrix, we refer
 to \cite{Glauber69,Nussenzveig}.
 }
 \ba
   \Delta_v w^2 = \langle \abs{w}^2 \rangle - \abs{\bar w(v)}^2
   = 1+\delta \, .
 \ea
 In the case of two-mode squeezed states ($\delta = 0$),
 this is exactly $1$. (Thus the second term of the exponent is the overlap
 of coherent states
 $\abs{\langle w \vert \bar w \rangle}^2 = e^{-\abs{w-\bar w}^2}$,
 and one recovers again the pure state
 Eqs. (\ref{insuperpcoh}, \ref{vwdistrib})).
 For the
 decohered  distribution \Eqref{rhodec}
 with parameter $\delta = 2$,
 one has $\Delta w^2 = 3$. This value should be compared with the
 corresponding width of the centered distribution in $v$,
 that is $\Delta v^2 = \av{\abs{v}^2} = n+1$.
 The ratio of the spreads is
 \ba
   \frac{\Delta w^2}{\Delta v^2} =
   \frac{1+\delta}{n+1}
   \, .
 \ea
 Therefore we obtain
 \ba \label{strongpeak}
   \frac{\Delta_v w^2}{\Delta v^2} \ll 1 \qquad
   \Longleftrightarrow \qquad \delta \ll n
   \, .
 \ea
 First, when it is verified, the conditional probability
 to find $w$ given $v$
 is sharply
 peaked around the conditional value $\bar w$. Therefore the modes
 ${\bf k}$ and $-{\bf k}$ are
 strongly correlated in phase and amplitude
 as discussed below \Eqref{vwdistrib}.
 Consequently, the power spectrum and the temporal coherence, see
 Eq. \Eqref{power}, are preserved whenever decoherence processes
 induce $\delta \ll n$.
 Put differently, whenever \Eqref{strongpeak} applies,
 these
 statistical properties of the distribution are as well
 described by statistical mixtures of entangled
 two-mode coherent states
 $\ket{v} \ket{\bar w}$, as in Eq. \Eqref{rhodec2}.

 Second, \Eqref{strongpeak}
 {\it is} the condition of factorizability of the two-point
 function Eq. \Eqref{neqqq}.
 What we shall learn is that this is a rather mild
 constraint which includes a wide range of density matrices.
 Indeed as we shall see below,
 this condition is satisfied
 by density matrices
 which have completely lost their quantum properties.

 \subsection{Macroscopic correlations}

 As discussed after Eq. \Eqref{vwrhov'w'},
 the pure squeezed vacuum state cannot be
 interpreted as a classical distribution because
 macroscopically separated
 ($n \geq \abs{v-v'}^2 \gg 1$)  semi-classical
 configurations are correlated.
 We now show that this property is very sensitive to the value of $\delta$.

 The width of these correlations is
 governed by
 the coefficient of the cross terms $v^*v',\, w^*w'$
 in Eq. \Eqref{offdiag}, i.e.
 $1 - (n+1)/\Delta  = 
 \delta/(\delta + 1)$.
 It belongs to the interval
 $\left[ 0 , \, n/(n+1) \right]$.
 The lower and upper bounds are reached for
 squeezed vacuum states and thermal states respectively.
 To better understand its physical meaning, it is
 instructive to pose and examine these two limiting cases. The
 former is presented in Section III.B.
 For a one-mode
 thermal state, one has
 \ba
   \abs{\bra{v} \hat \rho \ket{v'}} = \frac{1}{n+1} \exp\left[
   -\frac{1}{4}\left( 1+\frac{n}{n+1}\right)\abs{v-v'}^2 -
   \frac{1}{4(n+1)} \abs{v+v'}^2  \right]\, . \nonumber
 \ea
 The first term in the exponent is bounded by
 $1/2$, which means
 that the two different semi-classical configurations are
 uncorrelated. The second term is the power spectrum.



 We are now ready to interpret the results for \Eqref{offdiag}.
 To simplify the discussion,
 we calculate the width of the off-diagonal matrix elements
 \Eqref{offdiag} when
 $w=\bar w = cv^*/(n+1)$ and $w'=\bar w'=cv'^{*}/(n+1)$,
 i.e. for the most probable value of the conditional
 amplitudes of the $-{\bf k}$ mode.
 Then one has
 \ba \label{D12e}
   &&\abs{\bra{v}\bra{\bar w} \hat \rho \ket{v'}\ket{\bar w'}} =
   \frac{1}{\Delta} \exp\left[ -\frac{\abs{v-v'}^2}{2\sigma_{v}^2(\delta)} -
   \frac{\abs{v+v'}^2}{4(n+1)}  \right] \, , \nonumber \\
   && \frac{1}{\sigma_{v}^2(\delta)} = \frac{1}{2(n+1)} +
   \frac{\delta}{1+\delta}\left[ 2 -\frac{1+\delta}{n+1} \right]
  \, .
 \ea
 The width $\sigma_{v}$ on the line $v-v'$ is written as
 a function of $\delta$.
 Its inverse is plotted on Fig. $2$
 for a value of $n=100$. For large occupation numbers, $\sigma^2_{v}(\delta)$
 drops from its maximum value $2(n+1)$ (for the two-mode squeezed state)
 to $1$ over a range of
 $\delta = 1 + O(1/n)$.
 In conclusion,
 \ba
   {\rm{uncorrelated \,\,\, semiclassical \,\,\, states}}
   \quad \Longleftrightarrow \quad \sigma_{v}
   \leq 1   \quad \Longleftrightarrow \quad
   \delta \geq 1  \,  .
 \ea

 \begin{figure}[h] \label{macrospread}
 \vbox{ \hbox to\linewidth{\hss
    \resizebox{7.5cm}{6.5cm}{\includegraphics{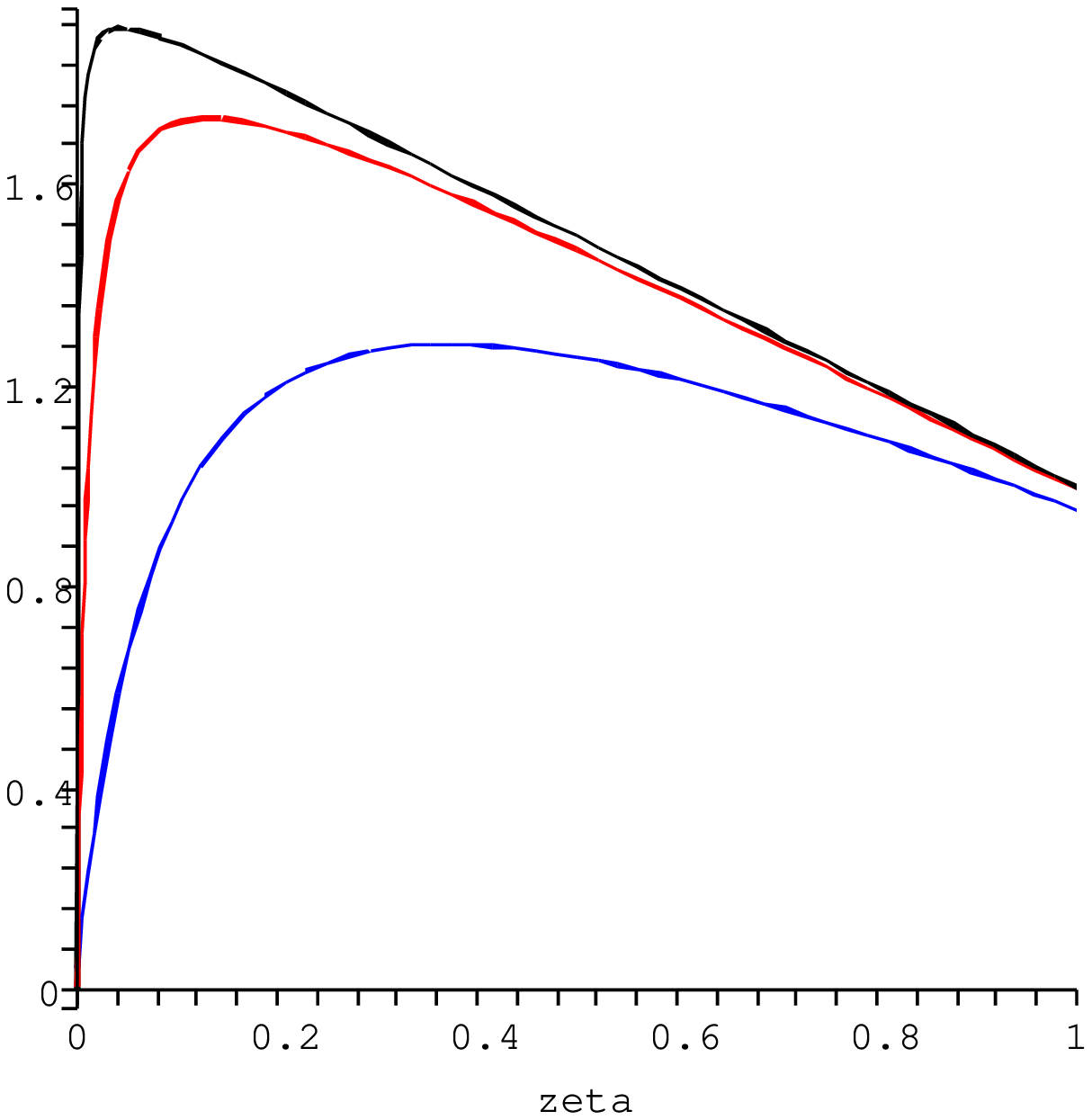}}
 \hspace{5mm}
        \resizebox{7.5cm}{6.5cm}{\includegraphics{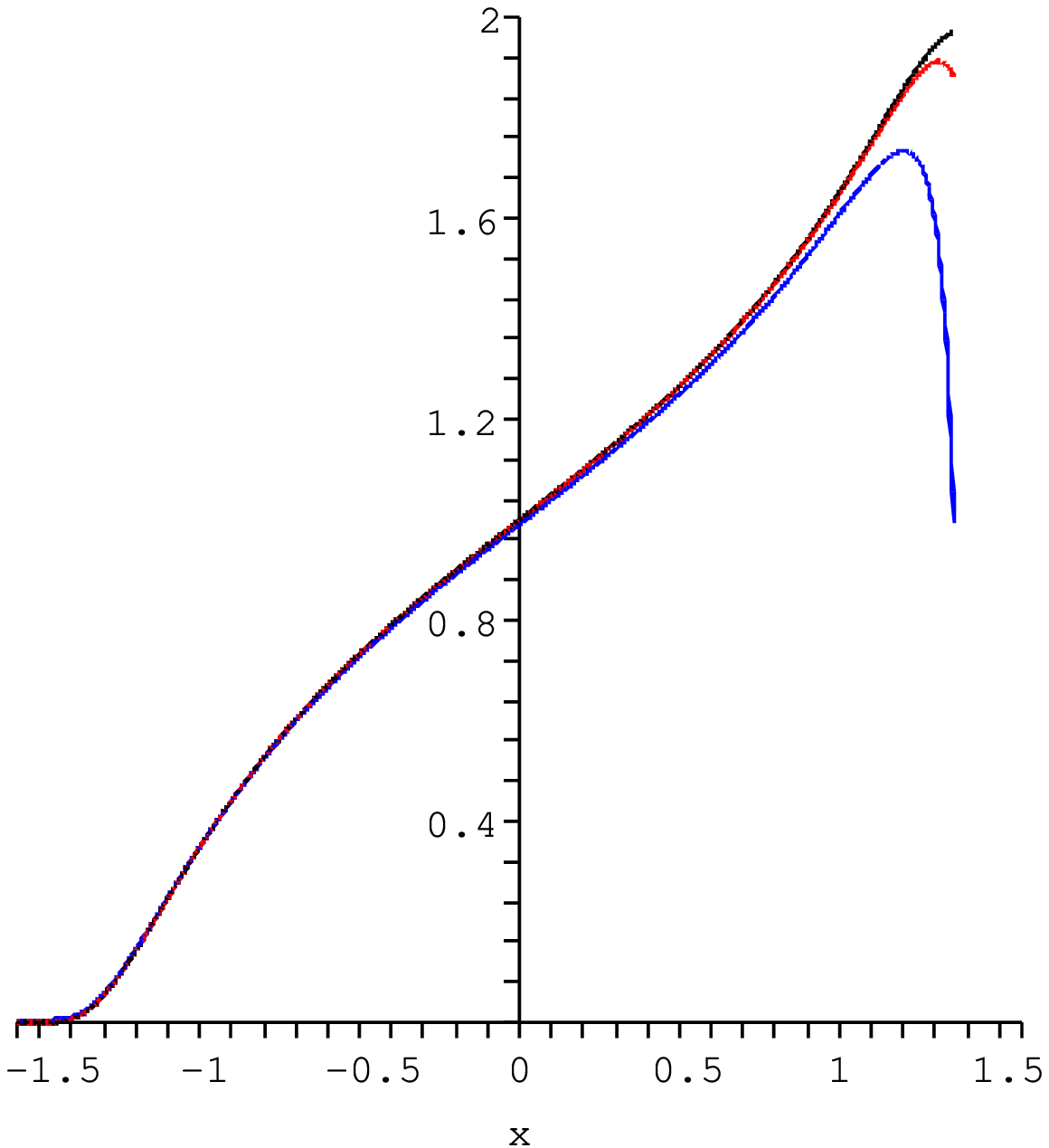}}}
\hss}
 \caption{{\it The inverse width $\sigma_{v}^{-2}$ of the off-diagonal
 matrix elements as a function of the degree of coherence of
 the distribution}
 for $n=10^1$ (lower, blue),
 $n=10^2$ (middle, red), and $n=10^3$ (upper, black).
 Left panel, as a function of
  $\zeta = \delta/n$,
 one clearly sees the extreme sensitivity of $\sigma_{v}$
 with respect to $\delta$ for small values.
 On the right panel, the same spread as
 function of $x=\arctan(\ln(\delta))$.
 This further shows that, 
 for {\it all} values of $n \gg 1$, the critical regime is
 centered around $\delta = 1$,
  i.e. the `grain' of
  coherent states.}
\end{figure}


 \subsection{Sub-fluctuant mode and the power of the decaying mode}

 From the second equation in \Eqref{eigenvalues},
 we see that the sub-fluctuant mode behaves as
 \ba  \label{RRR}
   \langle \{ \Pi_{\bf k} , \, \Pi^\dagger_{\bf k} \} \rangle
   =  \left( \frac{\delta}{2} + \frac{(1+\delta)^2}{8n} + O(n^{-2})
   \right)
   \, .
 \ea
 Therefore one looses the squeezed spread,
 i.e. $\langle \{ \Pi_{\bf k} , \, \Pi^\dagger_{\bf k} \} \rangle \geq 1/2$,
  when $\delta \geq 1$. In conclusion, one has
 \ba
   {\rm{no \,\,\, subfluctuant \,\,\, variable}} \quad
   \Longleftrightarrow \quad \delta \geq 1  \, .
 \ea

 It is interesting to notice that the
 value
 of the quantum coherence $\sigma_v$ and that of the squeezed spread
 are correlated.
 Indeed, their product obeys
 \ba \label{interprod}
   \sigma_v^2 \,
    \langle \{ \Pi_{\bf k} , \, \Pi^\dagger_{\bf k} \} \rangle
   = \frac{1+\delta}{4}\left(1 + O(\frac{1}{n}, \frac{\delta}{n})\right) \, ,
 \ea
 for  $0 \leq \delta \ll n$. (This result is non-trivial for $\delta \leq 1/n$.)
 In Fig. 3, we have represented the plot of this product for the whole range of
 $\delta$.
 It           is growing with $\delta$, the
 lower bound $1/4$ being reached for pure (squeezed) states.
 More importantly it           stays
 almost constant
 for $\delta \leq 1$ before growing linearly with $\delta$ for $\delta \gg 1$.

 The first regime corresponds to the loss of the quantum coherence
 whereas the second corresponds to the linear growth of the power
 of the subfluctuant mode
 (since $ \sigma_v^2 $ stays $\simeq  1$).
At this point it is important to notice that this second regime
also corresponds to a linear growth of the power
 of the decaying mode. Indeed, the powers of
 the subfluctuant mode and the decaying mode are related by
 \ba  \label{RRR2}
   \langle \{ d_{\bf k} , \, d^\dagger_{{\bf k}} \} \rangle
   &=&\langle \{ \Pi_{\bf k} , \, \Pi^\dagger_{\bf k} \} \rangle
  + \vert c \vert (1 - \cos 2 \theta)
   \nonumber \\
   &=&  \langle \{ \Pi_{\bf k} , \, \Pi^\dagger_{\bf k} \} \rangle
  + O(n^{-1/2})
 = \frac{\delta}{2} + O(n^{-1/2})
   \, ,
 \ea
 where $\theta$
 is defined in Eq.
 \Eqref{epsilontheta}.
 In the second line, we have used the inflationary value of
 $\theta$ ($\propto n^{-3/4}$)
 and the fact that $\theta$ is unchanged when increasing $\delta$.

Similarly, the cross-correlation is
 \ba  \label{RRR3}
   \langle \{ d_{\bf k} , \, g^\dagger_{{\bf k}} \} \rangle
    = - 
    \sin(2\theta) \vert c \vert \,
    \propto n^{1/4} \left( 1 + \frac{1-\delta}{2n} + O(n^{-2}) \right)
   \, .
 \ea
 One sees that the cross term is unaffected as long as
 $\delta \ll n$.

 \begin{figure}[ht]
 \epsfxsize=10.0truecm
 \epsfysize=10.0truecm
 \centerline{{\epsfbox{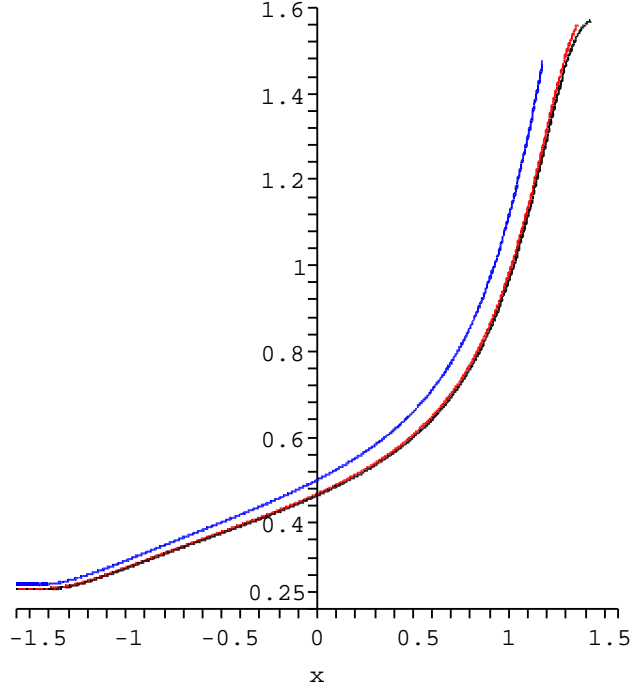}}}
 \caption{{\it The product of the width $\sigma_v^2$ with the
 power  of the subfluctuant mode},
 for $n=10^1$ (upper, blue),
 $n=10^2$ (middle, red), and $n=10^3$ (lower, black).
 For the horizontal axis, we used the non-linear scale
 $x=\arctan(\ln(\delta))$ of Fig 2. For
 the vertical axis, we used the non-linear scale
 $y=\arctan(\sigma_v^2 \langle \{ \Pi_{\bf k} , \, \Pi_{-\bf k} \}
 \rangle)$. 
 One clearly sees that 
 for all $n \gg 1$, the product of Eq. \Eqref{interprod}
 stays between $1/4$ and $1/2$ as long
 as $\delta \leq 1$.
 One also verifies that it is asymptotically $n$-independent
 for all values of $\delta \ll n$.
 Finally we recall that the power of the decaying mode
 coincides, to leading order, with
 $\langle \{ \Pi_{\bf k} , \, \Pi_{-\bf k} \}
 \rangle $, see Eq. \Eqref{RRR2}.}
 \end{figure}

 \subsection{$P$-representability of partially
 decohered density matrices}

  We now discuss the
 notion of $P$-representability of a density matrix \cite{Glauber63b}.
 It is an intrinsic property of states which enlightens the
 distinct behaviours of quantum and classical distributions.

 A one-mode density matrix
 is said to be $P$-representable if it can be written as a
 mixture of coherent states
 \ba \label{defPrepr}
   \hat \rho_{P-\rm{repr}} = \int\!\!\frac{d^2v}{\pi}
   \, P(v) \, \ket{v,\, {\bf k}}\bra{v,\, {\bf k}} \, ,
 \ea
 where the weight function $P(v)$ is a
 tempered distribution \cite{Cahill65,Wodkiewiczlecture}.
 Therefore, when considering $P$-representable {\it Gaussian density matrices},
 the weight function $P(v)$ is simply a Gaussian function.
 The generalization of this definition
 to two-mode (Gaussian) distributions is straightforward:
 \ba \label{twomodeP}
   \hat \rho_{P-\rm{repr}} = \int\!\!\frac{d^2v}{\pi}
   \frac{d^2w}{\pi}
   \, P(v,w) \, \ket{v,\, {\bf k}}\bra{v,\, {\bf k}}
   \otimes
   \ket{w,\, -{\bf k}}\bra{w,\, -{\bf k}}
   \, .
 \ea
 The $P$-representation of the distribution
 characterized by Eqs. \Eqref{momenta} can easily be calculated.
 From \Eqref{twomodeP}, one calculates the expectation values
 ${\rm Tr}(\hat n_k \hat \rho_{P-\rm{repr}}) = n$ and
 ${\rm Tr}(\hat a_{\bf k} \hat a_{-{\bf k}} \hat \rho_{P-\rm{repr}}) =
 c$, and one gets
 \ba \label{Puvp}
   P(v,w) = \frac{1}{\Delta'} \exp\left[-\frac{\abs{v}^2}{n}
   - \frac{n}{\Delta'} \abs{w - (cv^*/n)}^2 \right] \, ,
 \ea
 where $\Delta' = n^2 - \abs{c}^2$ must be positive.
 The function $P$ is well defined if,
 and only if, $n \geq \abs{c}$,
 i.e. $\delta \geq n/(n+1)$.
 Hence, in the high occupation number limit,
 \ba \label{Prepr}
   {\rm{P-representability}}
   \quad \Longleftrightarrow
   \quad \delta \geq 1 \, .
 \ea
 This condition
 coincides with that for the
 non-existence of a sub-fluctuant variable,
 see Eq. \Eqref{RRR}, as well as that for the vanishing of
 the off-diagonal matrix elements, see Eq. \Eqref{D12e}.
 It also coincides with the condition obtained from the criterion
 of separability we shall discuss in the next subsection,
 see Eq. \Eqref{separability}.
 This proves that {\it all} quantum properties are lost 
 if, and only if, $\delta \geq 1$.

 It is therefore of interest to further analyze this case
 and to determine how it is related  to two-mode coherent states.
 Consider the limiting case $\Delta' \to 0^+$,
 i.e. $\delta \to n/(n+1) $ from above.
 In this limit $P(v,w)$ of Eq. \Eqref{Puvp} becomes
 \ba
 \label{lim}
   P_{lim}(v,w) =
   \frac{1}{n} \exp\left[-\frac{\abs{v}^2}{n} \right] \,
   \delta^{(2)}\left(w - \frac{c}{n}v^* \right) \, .
 \ea
The presence of the two Dirac functions constraining
the real and imaginary parts of $w$
clearly expresses the critical character of this distribution.

One can now notice that this distribution corresponds to
the simplified distribution of Eq. \Eqref{rhodec2}.
 %
 More precisely, the equality of the entropies, see Appendix D.7,
 suggest that \Eqref{lim} is the quantum counterpart of the
 classical effective classical distribution
 Eqs. (\ref{sineeq}-\ref{Peff})
 obtained by setting the decaying mode to zero.
 The (badly named) {\it minimally decohered} distribution
 \Eqref{rhodec} is in fact slightly more decohered as can be
 understood from the replacement of the Dirac's
 $\delta^{(2)}\left(w - \frac{c}{n}v^* \right)$ by a Gaussian with
 a width equal to 1. This subtle difference is better appreciated
 when one realizes that the simplified distribution of Eq.
 \Eqref{rhodec2} and that of \Eqref{rhodec},  which are both
 diagonal in two-mode coherent states, belong to the {\it same
 class} of decohered matrices when considering the entropy.


 \subsection{Separability of two-mode homogeneous distributions
 and violations of Bell's inequalities}

 The non-separability
 is another important criterion for distinguishing
distributions which cannot be viewed as classical stochastic distributions.
Its physical relevance comes from the fact that the non-separability
is a necessary property to have expectation values violating
Bell's inequalities\cite{Werner89}.
At the end of this subsection, we have included
a brief discussion concerning the relationship
between decoherence and the associated loss of
the violation of Bell's inequalities.
For more details we refer to \cite{CP4}.

 A word of caution is in order.  Unlike the
 $P$-representability
 the non-separability rests on a division of the
 $4$ canonical variables into two subsets of $2$ canonical variables.
 To define the two subsets, we use creation and destruction operators
 carrying a momentum ${\bf k}$ or ${\bf -k}$. The criterion should
 therefore be called the  $({\bf k}, {\bf -k})$-separability.
 The justification of the choice of the two subsets
 is presented after having explained the concept of separability.

 Following Werner \cite{Werner89},
 a distribution of two-modes
 $\bf k$ and $-{\bf k}$ is said to be
 separable if, and only if,
 it can be written as a convex sum of
 products of distributions, i.e.
 \ba \label{Defseparability}
   \hat \rho_{sep} = \sum_{l} p_l \, \hat \rho_{\bf k}^{(l)} \otimes
   \hat \rho_{-{\bf k}}^{(l)} \, , \qquad p_l \geq 0 \, , \qquad
   \sum_l p_l = 1 \, .
 \ea
 Notice that the definition \Eqref{Defseparability}
 is not the statement that a separable
 state is
 a product of
 its reduced density matrices
 (in which case one would obtain uncorrelated sub-systems:
 $p_l = 0$ for all $l$ but one).

 The key point is that
  all states of the form
 \Eqref{Defseparability} can be obtained by
 the following classical protocol: when a random generator
 produces the number $l$ with probability $p_l$,
 two distant observers performing separate
 measurements on the subsystems
 $\bf k$ and $-{\bf k}$ respectively, prepare them
 into the states $\hat \rho_{\pm {\bf k}}^{(l)}$.
 Separable states are said classically correlated because, by construction,
 their statistical properties can be interpreted
 classically. In particular they cannot violate Bell's inequalities.
 States which are not separable are called ``entangled''
 (such as a two-mode squeezed vacuum states, or a singlet
 of $1/2$ spins) and can only be produced by letting the two
 parts of the system interact.

 Although it may be very difficult to decipher the separability of
 a general two-mode state,
 there is a criterion for Gaussian
 states. We briefly outline the proof of this theorem,
 the complete proof can be found in \cite{Simon00}.
 First, one notices on \Eqref{Defseparability} that
 the
 separability is a property
 that is not affected by
 {\it local} unitary transformations
 $U_{loc} = U_{{\bf k}} \otimes \tilde U_{-{\bf k}}$.
 These local
 transformations are of two types, 
 rotations and squeezing 
 which do not mix $\bf k$ and $\bf -k$.
 Therefore,
 if $\hat \rho$ is a separable state, then all states
 $\hat \rho\, '= U_{loc} \, \hat \rho \, U_{loc}$
  are also separable.
 Second, one notices that if $\hat \rho$ is a $P$-representable state
 \Eqref{twomodeP}, it is separable. The non trivial part of the
 proof consists then to show that {\it all} Gaussian separable
 states can be constructed by applying local unitary transformations
 on a $P$-representable state.
 In brief, the criterion for separability for two-mode
 Gaussian states is
 {\it a Gaussian state $\hat \rho$ is separable if, and only if
 there is a local unitary
 transformation $U_{loc}$ which brings it to a $P$-representable
 state}.

 These local transformations are very constrained for  homogeneous
 states because any local squeezing would break the homogeneity of the state
 (one would have either $\langle \hat a_{\bf k}^2 \rangle \neq 0$
 or $\langle \hat a_{\bf  -k}^2 \rangle \neq 0$,
 or both).
 Hence, $U_{loc}$ can only be a product of local rotations.
 These preserve
 the non-negativity of the Gaussian function $P(v,w)$.
 As a result, separability and
 $P$-representability are equivalent properties of
 homogeneous Gaussian states.
 Given \Eqref{Prepr}, one concludes 
  \ba \label{separability}
   ({\bf k}, -{\bf k})-{\rm{{separability
   \,\,\, of \,\,\, homogeneous\,\,\, states}}}
   \quad \Longleftrightarrow
   \quad \delta \geq 1 \, .
 \ea

 Notice that the $({\bf k}, {\bf -k})$-separability and
 the $(\phi_1 , \phi_2)$-separability
 are two independent properties of the state.
 The former 
 implies the above inequality on the decoherence level.
 As shown in Appendix C, the latter is simply a consequence of
 homogeneity of the distribution.
 The independence of these two concepts can be understood by
 noticing that the separability
 defines an equivalent class of density matrices
 through {\it local} transformations only.
 In fact,
 the transformation from the operators with a definite
 momentum ${\bf k}$ to ($\phi_1$, $\phi_2$) is a global, $4$ by $4$,
 transformation, see \Eqref{U}.

 We should now explain the selection of the two subsets.
 The choice of $(\hat a_{\bf k}, \, \hat a^\dagger_{\bf k})$ and
 $(\hat a_{\bf -k}, \, \hat a^\dagger_{\bf -k})$, and not
 $(\hat a_{1}, \, \hat a^\dagger_{1})$ and $(\hat a_{2}, \, \hat
 a^\dagger_{2})$, is dictated by the
 nature of the interactions amongst modes. When keeping non-linear terms,
 the Lagrangian density is a sum of products of
 fields evaluated at the same point.
 This leads to terms of the form
 $\phi_{{\bf k}} \phi_{{\bf p}} \phi_{-{\bf k}-{\bf p}}$
 which are governed by the Fourier
 transform of the field $\phi(\eta,{\bf x})$. These non-linear terms
 give rise to momentum exchanges
 which  cannot be described by $\hat a_1$ or $\hat a_2$ separately.
 (It should be noticed that these hermitian operators obey unusual commutation
 relations with the momentum operator $\hat {\bf P}$:
 $[\hat {\bf P},\, \hat a_{1,\,2}] = \mp i {\bf k} \,  \hat a_{2,\,1} $).

 Notice that contrary to non-linear interactions, bilinear
 couplings do not permit to 
 dismiss $\phi_1, \, \phi_2$.
 %
 %
 %
 %
 Indeed, consider a  bilinear
 interaction with a random field by adding a coupling term
 $\int\!\!d^3x \, \hat \phi(\eta,{\bf x}) \, \Psi (\eta,{\bf x})$
 where $\Psi (\eta,{\bf x})$ is a real stochastic field.
 Then the preferred basis
 are still the real and imaginary parts of $\phi(\eta,{\bf x})$,
 see \cite{KiefPolStar98a}.
 To verify this it suffices to
 decompose the interaction term in Fourier transform, use the
 reality condition of the stochastic field $\Psi (\eta,{\bf x})$
to get
$\int\!\!\widetilde{d^3k} \, \left[\hat \phi_1(\eta, {\bf k}) \, \Psi_1(\eta, {\bf k})
 + \hat \phi_2(\eta, {\bf k}) \, \Psi_2(\eta, {\bf k})\right]$,
and then use the homogeneous and isotropic character
of the moments of $\Psi (\eta,{\bf x})$.
In this we recover the fact that
 homogeneity implies that all Gaussian density matrices
 are $(\phi_1,\, \phi_2)$-factorizable.

 It is nevertheless possible to consider observables
 which cannot be written in terms of $\hat a_1$ or $\hat a_2$ separately
 without introducing explicitely non-linear interactions. 
 To this end one should exploit the cosmic variance,
 namely the fact we do not observe the mean value, but rather
 some realization of the ensemble.
 (We refer  to Chap. V of
 \cite{CP04} for the procedure to extract some realization
 from the distribution, to Chap. VI for
 the calculation of observables associated with a given realization,
 and to the references of this work for the relationship
 between the role played by the interaction Hamiltonian in non-linear
 theories and the role of
 the projector introduced to extract some configurations
 from the mean.)
 In particular, upon specifying that a localized classical field
 configuration possessing a definite {\it velocity} (e.g. a
 localized wave packet made with running waves)
 has been realized/detected, the conditional
 expectation value of $\hat \phi(\eta, {\bf x})$ describes a pair
 of wave packets moving away from each other
 and which are spatially well separated.
 These wave packets localized both in ${\bf x}$ and ${\bf k}$ space
 are built with $\hat a_{\bf k}$ and $\hat a_{-{\bf k}}$ separately. Hence
 they cannot be written in terms of $\hat a_1$ or $\hat a_2$ separately.


\smallskip

{ \bf{Loss of violations of Bell inequalities}}

\noindent
In \cite{Wodkievicz}, the authors point out that
the projectors on a given field configuration
can violate
Bell's inequalities when evaluated in two-mode squeezed states.
In \cite{CP4},
we extend their analysis by replacing the pure states
by the partially decohered distributions of Eq. \Eqref{Qrepr}.
We plot in Fig. $4$ the combination $(n+1)\cal C$ of Eq. ($11$)
of \cite{CP4}
as a function of the square field amplitude
for $n=100$ and for three different values of $\delta$.
The points over the horizontal line
$(n+1){\cal C}=2$ give combinations of operators violating Bell's inequalities.
We observe that for $\delta > 0.1$, there is no more violations.
However,
the combination $\cal C$ does not provide the
operator which maximizes the violation.
Hence, there might still be operators which violate Bell's inequalities for
$0.1 \leq \delta < 1$.

\begin{figure}[h]
    \resizebox{7.5cm}{6.5cm}{\includegraphics{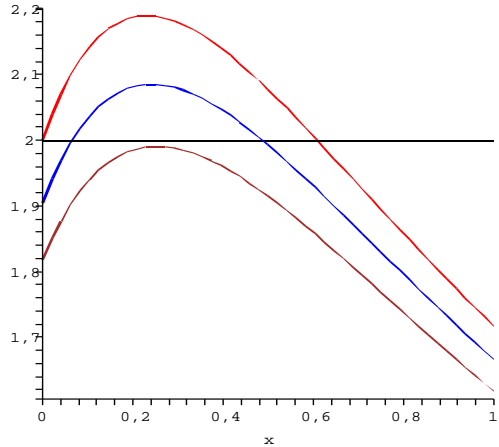}}
 \caption{{\it The loss of violation as decoherence increases.}
 The function is normalized to its maximal classical value $=2$.
 The variable $x$ gives the square field amplitude.
 The occupation number is $n=100$, and the three values of $\delta$
 are $0$ (red, upper), the pure state, $0.05$ (blue, middle), and
 $0.1$ (brown, lower), the border regime. It is also worth to point out that
 the function $(n+1){\cal C}(x)$ is independent of $n$ in the large $n$ limit
 as what is found in Figs. 3, 5, and 6.}
 \end{figure}

  \subsection{Entropy of partially decohered distributions}

 \begin{figure}[h] \label{macrospread}
 \vbox{ \hbox to\linewidth{\hss
    \resizebox{7.5cm}{6.5cm}{\includegraphics{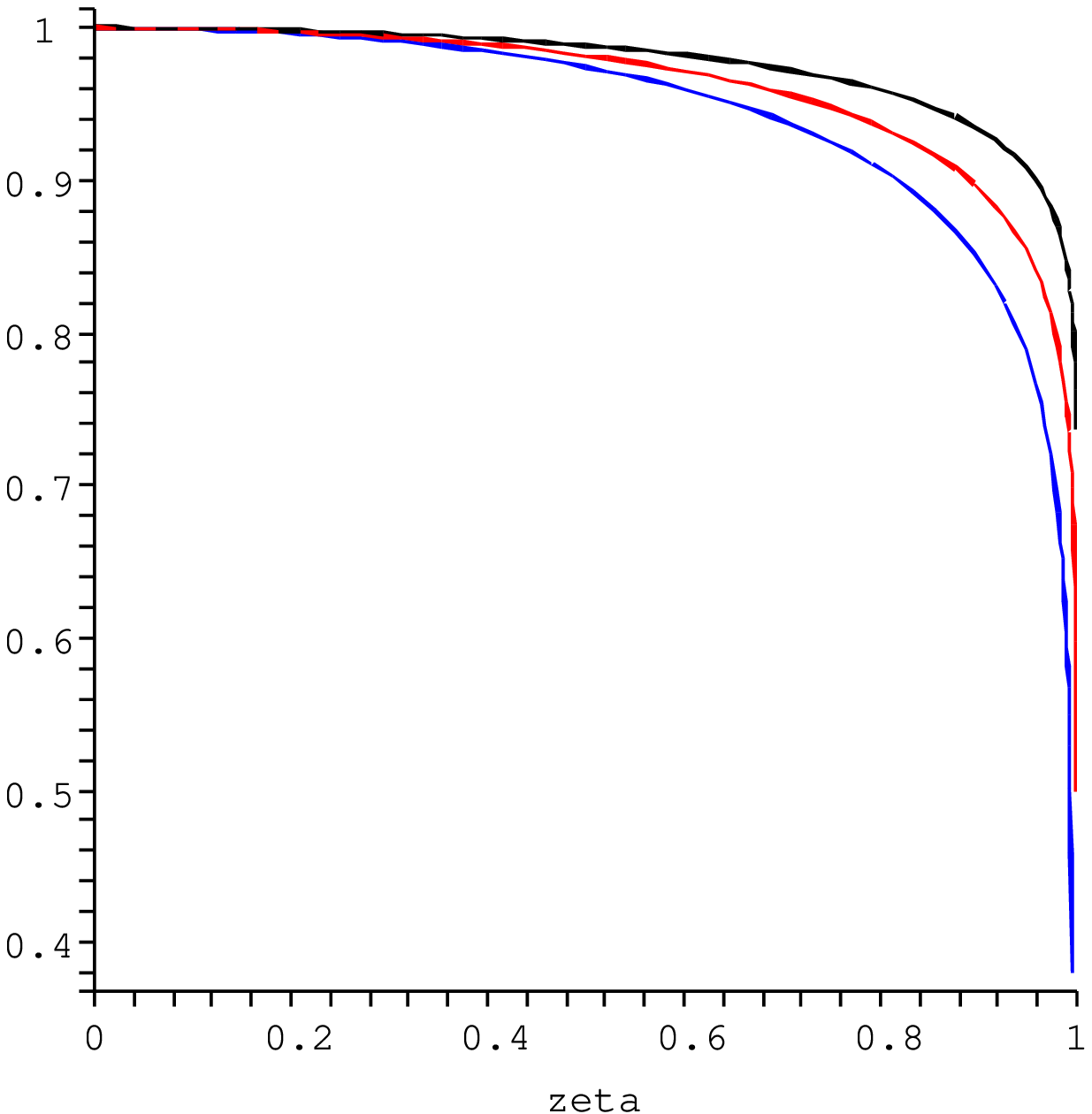}}
 \hspace{5mm}
        \resizebox{7.5cm}{6.5cm}{\includegraphics{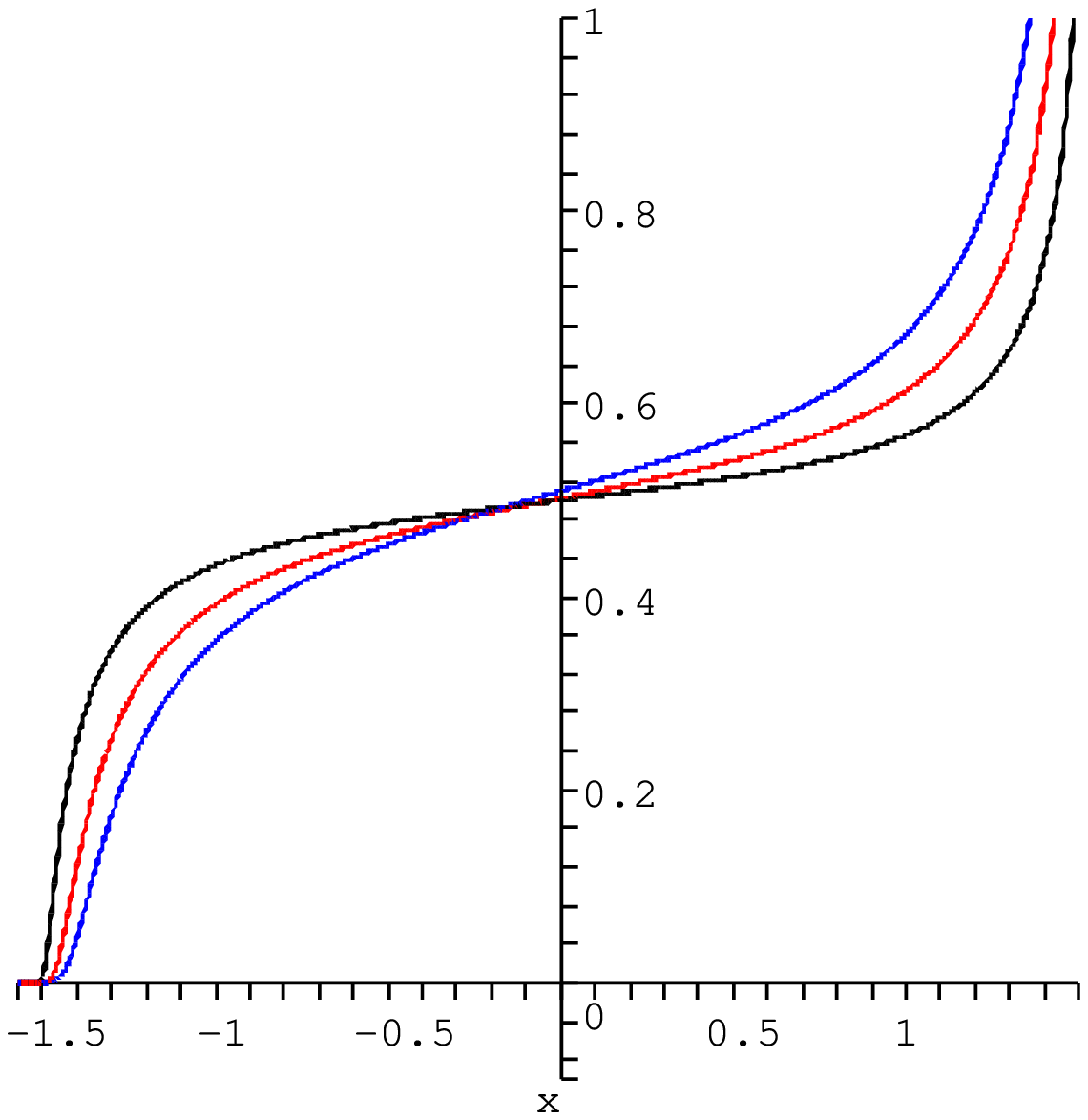}}}
\hss}
 \caption{{\it Entropy as a function degree of coherence
 for three values of $n$.} The function is normalized to it maximal value.
 The occupation number is $n=10^2$ (lower, blue),
 $n=10^3$ (middle, red), and $n=10^5$ (upper, black).
 On the left panel, as a function of the ratio
 $\zeta = \abs{c}/(n+1/2)$, 
 one clearly see the sensitivity for 
 $\zeta \to 1^{-}$. On the right panel, as a
 function of $x=\arctan(\ln(\delta))$, one verifies
 that the critical value $S_{max}/2$ 
 corresponds to
 $\delta =1$ for all $n$. 
 }
 \end{figure}

 To characterize the entropy of the general distribution
 \Eqref{Qrepr},
 we write $\delta = A n^\gamma,\, A> 0, \,  1 \geq \gamma \geq 0$.
 As we shall see, in the high occupation number limit,
 $\gamma$ is tightly constrained.

 The thermal occupation number defined
 in Eq. \Eqref{tilde n} and the entropy respectively scale as
 \ba
   \bar n = \sqrt{A} n^{(1+\gamma)/2} + O(n^{-(1+\gamma)/2}) \, ,
 \ea
 and
 \ba \label{paramd2}
   S^{(\gamma, A)}_{{\bf k},\, -{\bf k}} =
   (1+\gamma) 2 r + \left( 2 + \ln A \right)
   + O(n^{-(1+\gamma)/2}) \, .
 \ea
 We have used the squeezing parameter $n = \sh^2(r)$.
 From this analysis we see that the uncertainty in the definition of the
 quantum distributions which gives rise to the entropy of the effective
 distribution of sine functions of \Eqref{Peff}
 is very limited.
 (The entropy of that distribution is $S_{{\bf k},\, -{\bf k}}= 2 r$,
 see Eq. \Eqref{Smin} and the discussion in the following paragraphs.)
 Indeed, the linear dependence of the entropy in $r$
 implies that the distributions with entropy $= 2 r$
 all have
 \ba
  \gamma \ll \frac{1}{\ln n} \, .
 \ea

 When refining the analysis by taking into account the term in $\ln A$,
 and requiring that $\vert S_{{\bf k},\, -{\bf k}} - 2(r+1) \vert \leq B$,
 one gets
 $\vert \ln \delta  \vert \simeq \delta -1 \leq B$,
 independently of $n$.
 In the large $n$ limit, this
 fixes the value of the coherence term $c$
 with a precision $B/n$,
 thereby tightly restricting the quantum density matrices which
 can be put in correspondence with  Eq. \Eqref{Peff}.

Nevertheless, the criterium  $S_{{\bf k},\, -{\bf k}} = 2 r + O(1)$ 
does not discriminate between the
 distribution of Eq. \Eqref{rhodec} ($\delta = 2$)
 and that of  Eq. \Eqref{rhodec2} ($\delta = 1$).
 Hence the distribution \Eqref{rhodec} should be conceived
 as {\it minimal} only as 
 being a member of the {\it equivalent class
 of minimally decohered matrices}, i.e.
 those which have lost their quantum properties ($\delta > 1$)
 and which possess an entropy equal to $2 r$.

 \begin{figure}[h] \label{macrospread}
 \vbox{ \hbox to\linewidth{\hss
    \resizebox{7.5cm}{6.5cm}{\includegraphics{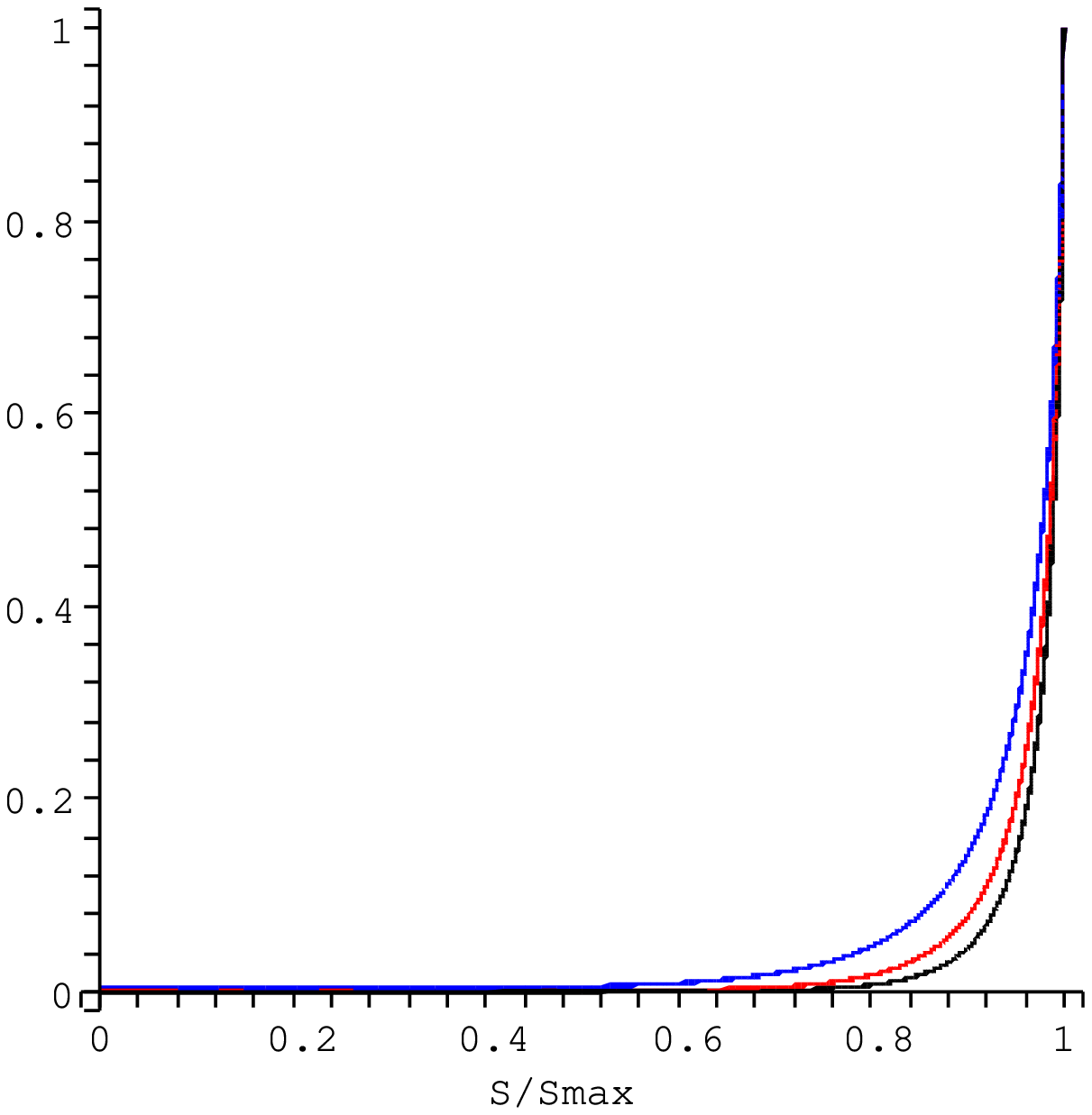}}
 \hspace{5mm}
        \resizebox{7.5cm}{6.5cm}{\includegraphics{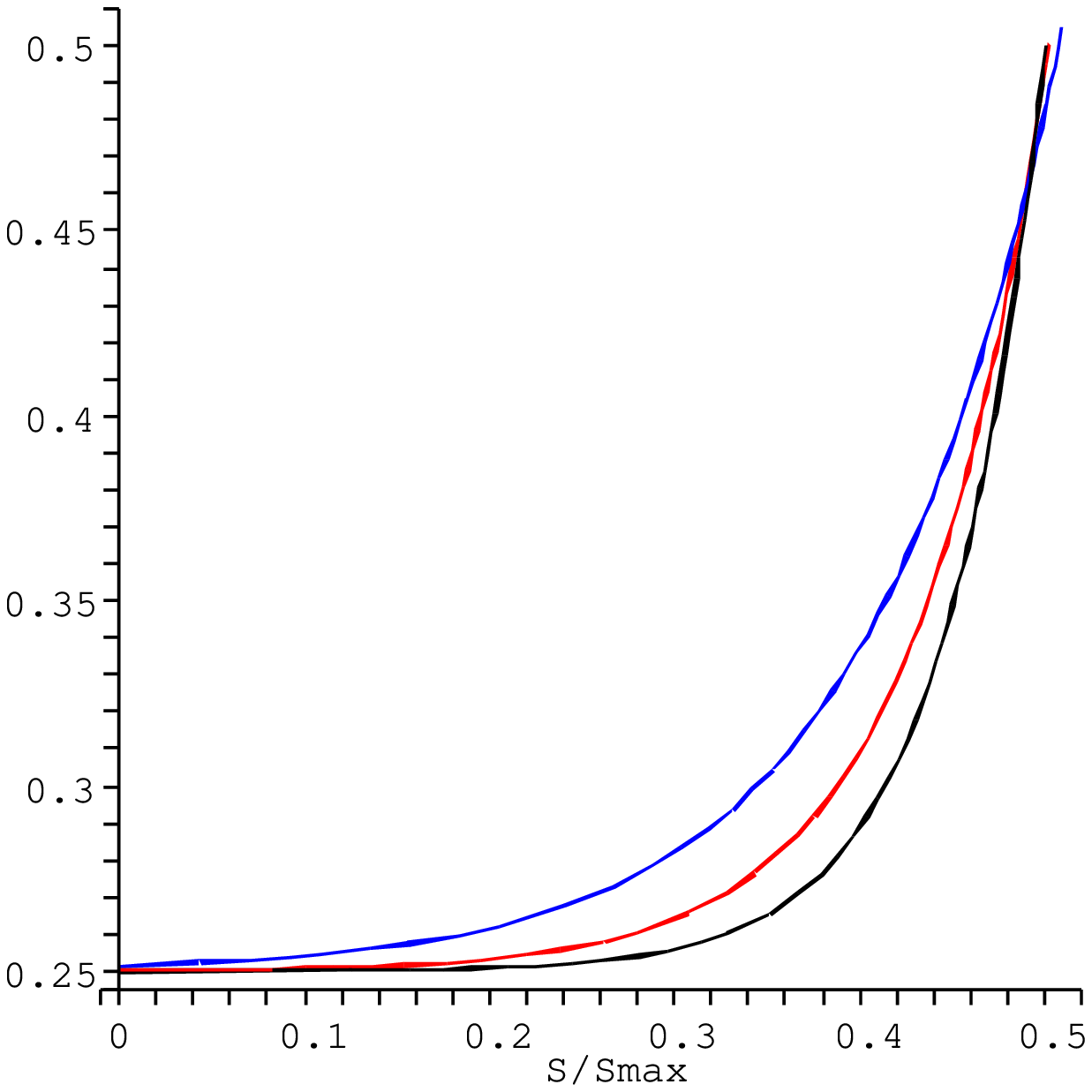}}}
\hss}
 \caption{{\it Parametric curves}
 We plot the product
 $\sigma^2_v \langle \{ \Pi_{\bf k} ,\, \Pi_{-{\bf k}} \} \rangle $
 versus the entropy. The latter (horizontal axis)
 has been normalized to its maximum value $S_{max}$.
 The occupation number is $n=10^2$ (upper, blue),
 $n=10^3$ (middle, red), and $n=10^5$ (lower, black).
 On the left panel, $\delta$ goes from $0$ to $n$ and
 the product
 $\sigma^2_v \langle \{ \Pi_{\bf k} ,\, \Pi_{-{\bf k}} \} \rangle $
 has been normalized to its maximum
 value $n$.
 On the right panel, we show a zoom on the region $\delta \leq 1$,
 and the product is not normalized.
 One sees the two regimes.
 First as long as the entropy has not reached
 half its maximum value, the product stays nearly constant.
 On the contrary, for $S > S_{max}/2$, the product explodes and reaches
 $n$. One can also admire
 the fact that the curves meet near $(1/2, 1/2)$ irrespectively
 of the value of $n$. 
 Moreover in the limit $n \to \infty$ the curve becomes an Heavyside
 function, thereby cleary separating the two regimes: the first one
 where quantum properties are progressively lost, and the second one
 where the power of the decaying mode linearly grows with $\delta$.}
 \end{figure}

 \subsection{Partial randomization of the phase}

 To complete the analysis we now
 assume that the phase of the squeezed vacuum
 state is randomly scattered.
  The density matrix
 $\hat \rho_{in}$ is then replaced by the statistical mixture
 \ba \label{rhophase}
   \hat \rho(r,\varphi_0) = \int_{0}^{2\pi}\!\!d\varphi \,
   f(\varphi;\, \varphi_0,\, \sigma) \hat \rho_{in}(r,\,\varphi) \, ,
 \ea
 where each pure state
 $\hat \rho_{in}(r,\,\varphi)$ is parameterized by the squeezing
 parameters $r$ and $\varphi$, and $f(\varphi;\, \varphi_0,\,
 \sigma)$ is the PDF of $\varphi$ centered at $\varphi_0$
 with a spread $\sigma$.
 We  assume that the mean phase shift vanishes. Therefore
 the mean value of $\varphi$ is the phase of the original distribution
 \Eqref{inoutvac}, namely $\varphi_0 = \pi + 2k\eta_r$.

 Since all the states in the ensemble \Eqref{rhophase} have the
 same value of $r$, the expectation value of the occupation number
 is still
 $\langle \hat a_{\pm {\bf k}}^{\dagger} \hat a_{\pm {\bf k}} \rangle =
 \sh^2 r = n$.
However the effective coherence term is reduced since it is affected
by 
 averaging over the phase :
 \ba
   \av{c}_{\varphi} =
   {\rm Tr}\left( \hat a_{{\bf k}} \hat a_{-{\bf k}}
   \hat \rho(r,\varphi_0) \right) =
   \sqrt{n(n+1)} \int_{0}^{2\pi}\!\!d\varphi \,
   f(\varphi) e^{i\varphi} \, .
 \ea
 Let us assume that the function $f(\varphi)$ is sharply peaked.
 Then we can extend the bounds of integration to infinity. Let us
 choose for instance a Gaussian PDF,
 $f(\varphi) = (2\pi\sigma^2)^{-1/2} \exp\left(
 -(\varphi - \varphi_0)^2/2\sigma^2\right)$. One gets
 \ba
   \av{c}_{\varphi} = \sqrt{n(n+1)} e^{i\varphi_0} e^{-\sigma^2/2}
   \, .
 \ea
 Hence, we have that $\abs{\av{c}_{\varphi}} =
 \abs{c_{sq}(r)} e^{-\sigma^2/2} <  \abs{c_{sq}(r)}$, i.e.
 the norm of the effective coherence term is smaller,
 as expected.
 A value $\delta_{cr} = 1$ corresponds to
 an extremely narrow distribution:  $\sigma_{cr}^2 = 1/(n+1)
 = 4(k\eta_r)^4 \sim 10^{-100}$.
 It should be noticed that these changes affecting the
 coherence are much smaller than those obtained
 when considering the trans-Planckian problem, see \cite{NP,NPC,MB},
 which concerned modifications of the power $n(k)$.


\end{appendix}

\end{document}